\documentclass[12pt,3p]{elsarticle}

\usepackage{amssymb}
\usepackage{amsthm}
\usepackage{amsmath}
\usepackage{siunitx}

\journal{Journal}

\begin{document}

\begin{frontmatter}

\title{Investigation on optimal microstructure of dual-phase steel with high strength and ductility by machine learning}

\author[1]{Misato Suzuki}
\author[2]{Kazuyuki Shizawa}
\author[2]{Mayu Muramatsu}

\affiliation[1]{organization={Department of Science for Open and Environmental Systems, Graduate School of Keio University},
            addressline={3-14-1, Hiyoshi, Kohoku-ku}, 
            city={Yokohama},
            postcode={223-8522}, 
            state={Kanagawa},
            country={Japan}}

\affiliation[2]{organization={Department of Mechanical Engineering, Keio University},
            addressline={3-14-1, Hiyoshi, Kohoku-ku}, 
            city={Yokohama},
            postcode={223-8522}, 
            state={Kanagawa},
            country={Japan}}

\begin{abstract}
In this study, we developed an inverse analysis framework that proposes a microstructure for dual-phase (DP) steel that exhibits high strength and ductility.
The inverse analysis method proposed in this study involves repeated random searches on a model that combines a generative adversarial network (GAN), which generates microstructures, and a convolutional neural network (CNN), which predicts the maximum stress and working limit strain from DP steel microstructures.
GAN was trained using images of DP steel microstructures generated by the phase-field method.
CNN was trained using images of DP steel microstructures, the maximum stress and the working limit strain calculated by the dislocation-crystal plasticity finite element method.
The constructed framework made an efficient search for microstructures possible because of a low-dimensional search space by a latent variable of GAN.
The multiple deformation modes were considered in this framework, which allowed the required microstructures to be explored under complex deformation modes.
A microstructure with a fine grain size was proposed by using the developed framework.

\end{abstract}

\begin{keyword}
Inverse analysis \sep Dual-phase steel \sep Convolutional neural network \sep Generative adversarial network \sep Phase-field method \sep Dislocation-crystal plasticity finite element method
\end{keyword}

\end{frontmatter}


\section{Introduction}
\label{sec1}
Dual-phase (DP) steel is composed of a soft phase (ferrite) and a hard phase (martensite).
The DP steel is widely used because of its significant mechanical properties.
However, the problem of DP steel is the trade-off between strength and ductility \cite{ritchie2011conflicts}.
For the material development of DP steel, it is necessary to find a microstructure with high strength and ductility \cite{tasan2015overview}.
Experiments have shown that the mechanical properties of DP steel are affected by the spatial distribution of martensite and ferrite.
For example, when the martensitic phase surrounds the ferrite phase to form a chainlike network structure, the strength of DP steel is higher and the ductility is lower than when the martensitic phase is isolated \cite{park2014effect}.
Other studies have shown that an increase in the proportion of martensite and the refinement of grains increase the strength, but ductility is less affected by grain refinement \cite{calcagnotto2011deformation,son2005ultrafine}.

However, material development requires repeated experiments through trial and error.
To reduce experimental costs, numerical simulations using computers are used in various fields \cite{TERADA20002285,BAI2020105735,WANG2023103948,YAP2023107771}.
For example, for steel, the phase-field method is used to predict the microstructure \cite{yeddu2012three,militzer2006three,TAKAHAMA20122916,MECOZZI2016245}, and the dislocation-crystal plasticity finite element method (FEM) is used to obtain mechanical properties \cite{woo2012stress,kadkhodapour2011micro,ZAAFARANI20061863,LU2020102703}.
These analysis methods are called forward analysis methods, but high computational costs become sometimes a problem.
Therefore, in recent years, methods of performing forward analysis rapidly and accurately have been explored \cite{LIAO2020105685,shinoda2004rapid,LOPES2021106650,NUTARO2023111990}.
In particular, machine learning is introduced for remarkably rapid and accurate forward analysis \cite{bock2019review,ramprasad2017machine,gubernatis2018machine,butler2018machine}.
For example, machine learning has been used for the prediction of material properties \cite{liu2019predicting,xie2018crystal,DUAN2013524} and the stress--strain relationship \cite{yamanaka2020deep,kalina2022automated}, and the improvement of processing \cite{gong2022additive,kouraytem2021modeling,RUIZ2022106785,zhu2021machine}.
Machine learning is also used for homogenization analysis, and it enables a more rapid computation of the mechanical properties of advanced materials, for example, porous and composite materials \cite{YANG2018278,lu2019data,krokos2022bayesian,li2019clustering,eidel2023deep}.
In DP steel research, machine learning is also used for various applications, including the prediction of properties \cite{marshall2021autonomous,li2019machine}, the optimization of processing \cite{khosravani2017development}, and the automation of phase segmentation \cite{MARTINEZOSTORMUJOF2022111638}.

In addition to forward analysis by numerical simulation, inverse analysis has also attracted attention as a means of reducing experimental costs in material development \cite{ronellenfitsch2019inverse,callewaert2016inverse,ZENG2023107920,liu2018generative,peurifoy2018nanophotonic}.
Inverse analysis proposes a material microstructure on the basis of the required mechanical properties, such as high strength and ductility.
Inverse analysis predicts in the opposite direction of forward analysis, which predicts mechanical properties from microstructures.
The microstructure proposed by inverse analysis provides an idea on the type of microstructure that should be made in experiments.
This enables a more efficient material development than the conventional trial-and-error process of material development without knowing the structure to be made.
Shiraiwa et al. \cite{shiraiwa2022exploration} performed inverse analysis to propose DP steel microstrustures.
Their study has the problem that the microstructure does not resemble the real one.
Hiraide et al. \cite{hiraide2021application} proposed a forward analysis method to predict the Young's modulus $E$ from a real polymer alloy phase separation structure and an inverse analysis method to output the structure from $E$.
This method does not have the aforementioned problem.
However, this method has not been applied to DP steel and that the investigation of the optimum microstructure is only performed within a specific deformation mode.
With the above background, the purpose of this study is to develop a machine learning model that proposes a DP steel microstructure exhibiting high strength and ductility.
The focus of this research is not to propose DP steel microstructures that are certain to be physically materialized, but to develop a framework for exploring DP steel microstructures that satisfy the required mechanical properties and have a high possibility to be materially embodied.
The inverse analysis framework proposed by Hiraide et al. \cite{hiraide2021application} is applied to DP steel and multiple deformation modes.

To apply the framework proposed by Hiraide et al. \cite{hiraide2021application} to DP steel, two main changes are made to the framework in addition to replacing the material.
The first is to modify the framework considering microstructures with high strength and ductility.
In the framework by Hiraide et al., it is necessary to specify the desired Young's modulus $E$ that microstructures should have.
In the material development of DP steels, it is necessary to consider the trade-off between strength and ductility, and to investigate a microstructure that is more compatible with both strength and ductility.
Therefore, in this study, we develop a framework considering a microstructure that maximizes the product of strength and ductility without specifically specifying the desired property values in advance.
The second is adapting the framework to four modes of deformation.
Whereas the framework developed by Hiraide et al. proposes a microstructure under one specified deformation mode, four deformation modes are considered simultaneously in this study: tensile toward the $x$ direction, tensile toward the $y$ direction, shear toward the $x$ direction, and shear toward the $y$ direction.
By considering multiple deformation modes, we can explore the microstructure required under complex deformation modes.

\section{Overview of inverse analysis framework}\label{sec2}
\begin{figure}[t]%
\centering
\includegraphics[width=0.6\textwidth]{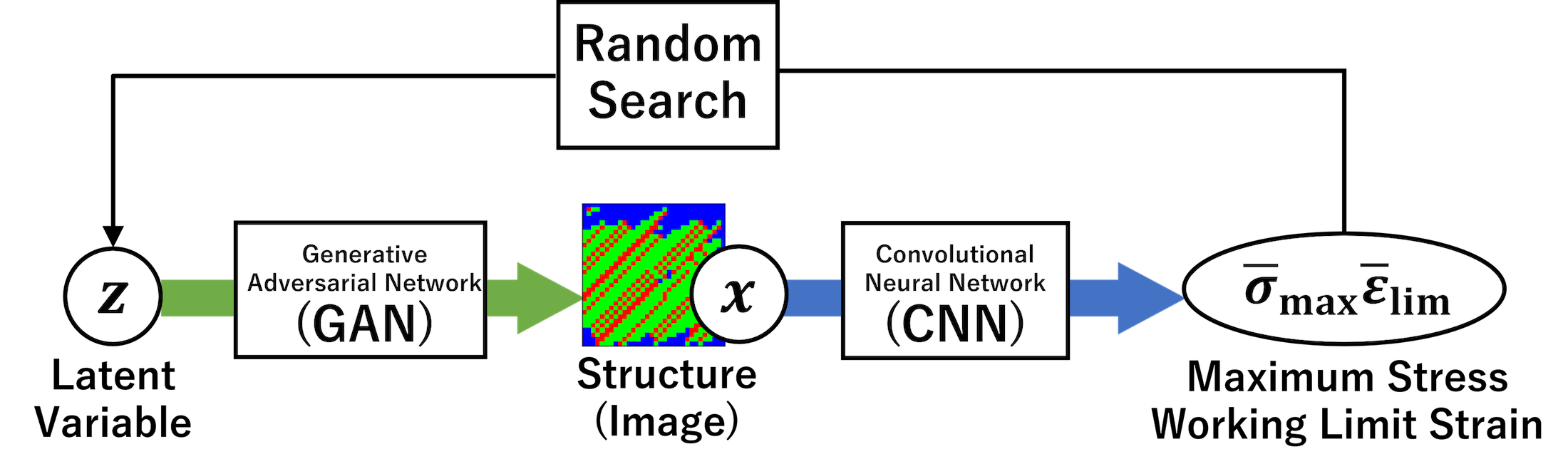}
\caption{Conceptual diagram of the inverse analysis framework applied to DP steel}\label{Fig1}
\end{figure}

A conceptual diagram of the inverse analysis framework is shown in Fig. \ref{Fig1}.
This inverse analysis framework employs two machine learning models: a generative adversarial network (GAN), which generates microstructures, and a convolutional neural network (CNN), which predicts mechanical properties from microstructures.

In this study, we apply the inverse analysis framework proposed by Hiraide et al. \cite{hiraide2021application} to DP steel.
Specifically, we use the maximum stress $\sigma_{\mathrm{max}}$ as the indicator for strength and the working limit strain $\varepsilon_{\mathrm{lim}}$ as the indicator for ductility.
To evaluate whether the microstructure has high strength and ductility, we use the product of the normalized maximum stress and working limit strain, $\bar{\sigma}_{\mathrm{max}}\bar{\varepsilon}_{\mathrm{lim}}$ \cite{adachi2020effect}.
The higher the value of $\bar{\sigma}_{\mathrm{max}}\bar{\varepsilon}_{\mathrm{lim}}$, the higher the strength and ductility of microstructures.

Inverse analysis involves repeated the random search on the model created by combining the trained GAN and CNN.
The random search is performed as follows.

\begin{description}
\setlength{\leftskip}{1.5em}
\item[\rm{Step 1:}] A series of latent variables $\boldsymbol{z}$ are randomly selected, and GAN outputs an image of a DP steel microstructure on the basis of the selected $\boldsymbol{z}$. \label{random_1}
\item[\rm{Step 2:}] From the image, CNN predicts the maximum stress $\sigma_{\mathrm{max}}$ and the working limit strain $\varepsilon_{\mathrm{lim}}$ for each of the four deformation modes: tensile toward the $x$ direction, tensile toward the $y$ direction, shear toward the $x$ direction, and shear toward the $y$ direction. \label{random_2}
\item[\rm{Step 3:}] For each deformation mode, the product of the normalized maximum stress and the working limit strain $\bar{\sigma}_{\mathrm{max}}\bar{\varepsilon}_{\mathrm{lim}}$ is calculated. \label{random_3}
\item[\rm{Step 4:}] When $\bar{\sigma}_{\mathrm{max}}\bar{\varepsilon}_{\mathrm{lim}}$ is the highest up to this step, the DP steel microstructure and deformation mode are saved as a tentative optimal solution. \label{random_4}
\item[\rm{Step 5:}] After sufficient iterations of the process from Steps 1 to 4, the microstructure and deformation mode of the DP steel with the maximum $\bar{\sigma}_{\mathrm{max}}\bar{\varepsilon}_{\mathrm{lim}}$ is determined as the optimal solution.
\end{description}

\section{Construction of GAN}\label{sec3}

In the inverse analysis framework shown in Fig. \ref{Fig1}, GAN is used to generate DP steel microstructures.
Training data for GAN are obtained by the phase-field method.
In section \ref{subsec1}, we describe the preparation of DP steel microstructures using the phase-field method, and in Section \ref{subsec2}, we describe the training of GAN.

\subsection{Phase-field method}\label{subsec1}
The phase-field method \cite{wang1997three} is used for predicting the evolution of material microstructures by using order parameters called phase-field variables.
In this case, we predict martensitic transformation.
In the martensitic transformation of steel materials, the lattice structure changes from a face-centered cubic (fcc) structure to a body-centered cubic (bcc) structure.
This change causes a compressive deformation in a particular axial direction.
For three-dimensional analysis, there are three directions in which the compressive deformation can occur: $x$-, $y$-, and $z$-axes.
Thus, the martensitic phases, which are crystallographically equivalent to each other although lattice deformation occurs in different directions, are referred to as variants.
The martensitic phases generated with the directions of compressive deformation toward the $x$-, $y$-, and $z$-axes are called variant$1$, variant$2$, and variant$3$, respectively.
Then, $\phi_{i}$($i=1,2,3$) is defined as the phase-field variables, which take $\phi_{i}=1$ in variant$i$ and $\phi_{i}=0$ in the other variants or in the matrix phase.
The time evolution equation for the phase-field variables $\phi_{i}$ is expressed by the following equation \cite{ALLEN19791085}:
\begin{equation}
\frac{\partial{\phi_{i}}}{\partial{t}}=-M_{\phi}\frac{\delta{\bar{G}}}{\delta{\phi_{i}}},
\label{eq1}
\end{equation}
where $M_{\phi}$ is the mobility of $\phi_{i}$ and $\bar{G}$ is the total free energy of the system expressed as
\begin{equation}
\bar{G}=\int_{V}(g_{\mathop{\mathrm{chem}}\nolimits}+g_{\mathop{\mathrm{grad}}\nolimits}+g_{\mathop{\mathrm{elast}}\nolimits})dV.
\label{eq2}
\end{equation}
Here, $g_{\mathop{\mathrm{chem}}\nolimits}$ is the chemical free energy density, $g_{\mathop{\mathrm{grad}}\nolimits}$ is the gradient energy density, and $g_{\mathop{\mathrm{elast}}\nolimits}$ is the elastic strain energy density.
The chemical free energy density $g_{\mathop{\mathrm{chem}}\nolimits}$ is expressed as the following equation to be metastable at $\phi_{i}=0$ and stable at $\phi_{i}=1$:
\begin{equation}
g_{\mathop{\mathrm{chem}}\nolimits}=\Delta{f}\left\{\frac{A}{2}\sum^{3}_{i=1}\phi^{2}_{i}+\frac{B}{3}\sum^{3}_{i=1}\phi^{3}_{i}+\frac{C}{4}\left(\sum^{3}_{i=1}\phi^{2}_{i}\right)^{2}\right\},
\label{eq3}
\end{equation}
where $\Delta{f}$ is the amount of change in chemical free energy during martensitic transformation and $A$, $B$, and $C$ are numerical constants.
The gradient energy density $g_{\mathop{\mathrm{grad}}\nolimits}$ is expressed as
\begin{equation}
g_{\mathop{\mathrm{grad}}\nolimits}=\frac{a^{2}}{2}|\nabla{\phi}|^{2},
\label{eq4}
\end{equation}
where $a$ is the gradient coefficient.
The elastic strain energy density is expressed as follows on the basis of phase-field microelasticity theory \cite{khachaturyan2008theory}:
\begin{equation}
g_{\mathop{\mathrm{elast}}\nolimits}=\frac{1}{2}\sigma_{ij}\varepsilon^{el}_{ij}=\frac{1}{2}C_{ijkl}\varepsilon^{el}_{kl}\varepsilon^{el}_{ij},
\label{eq5}
\end{equation}
where $\sigma_{ij}$ is the Cauchy stress and $\varepsilon^{el}_{ij}$ is the elastic strain.
In equation \eqref{eq5}, the Cauchy stress is expressed as $\sigma_{ij}=C_{ijkl}\varepsilon^{el}_{kl}$ using the elastic modulus $C_{ijkl}$.
By solving equation \eqref{eq1}, one can predict the spatial and temporal evolutions of martensitic transformation.

\begin{figure}[t]%
\centering
\includegraphics[width=0.6\textwidth]{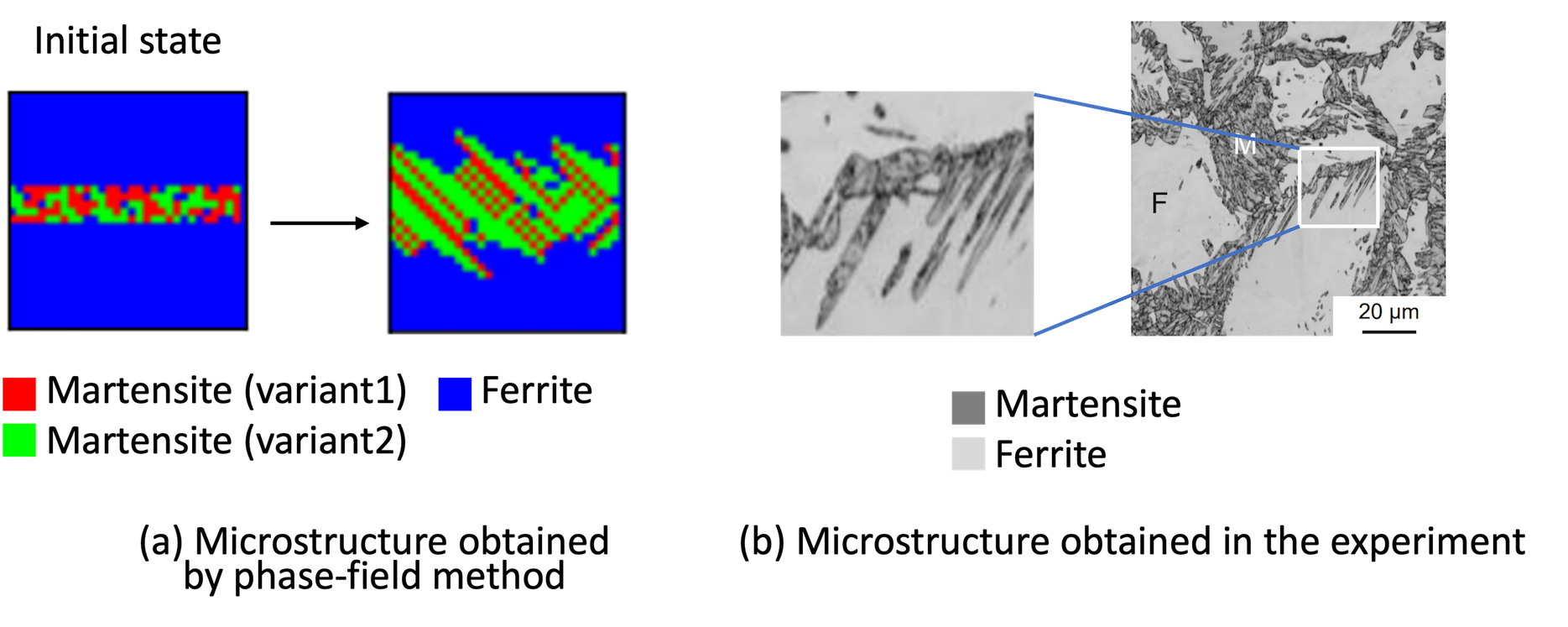}
\caption{An example of the initial conditions and generated microstructures in the phase-field analysis. (a) DP steel microstructure obtained by the phase-field method. (b) DP steel microstructure obtained in the experiment \cite{myeong2017effect}.}
\label{Fig2}
\end{figure}

The analytical conditions for the phase-field method are set as shown in Table \ref{Table1}.
The finite difference method is employed for the discretization.
Fig. \ref{Fig2}a shows an example of the initial conditions and generated microstructures in the phase-field analysis in this study.
By changing the crystallographic orientation of the matrix and the width of the initial grain boundary, we can generate various microstructures.
A prediction of martensitic transformation under these initial conditions corresponds to focusing on microstructures near the grain boundaries as shown in Fig. \ref{Fig2}b \cite{myeong2017effect}.
The generalization is therefore sufficient for microstructures near the grain boundaries.
For each of $170$ initial conditions, $10$ types of microstructure are output in a time series, generating a total of $1700$ microstructures.

\begin{table}[t]
\caption{Analysis conditions for phase-field method}\label{Table1}
\centering
\begin{tabular}{l l}
\hline
Dimensions of analysis & $2$ \\
Size &$\SI{31}{\um}\times\SI{31}{\um}$\\
Number of grid points &$32\times32$\\
Boundary condition & Periodic boundary condition \\
Change in chemical free energy $\Delta{f}$& $\SI{1.0}{kJmol^{-1}}$ \\
Elastic modulus $C_{11}$,$C_{44}$,$C_{12}$ & $\SI{397.0}{GPa}$,$\SI{123.5}{GPa}$,$\SI{150.0}{GPa}$ \\
Square of gradient coefficient $a^2$ & $5.0\times10^{-15}\si{\, Jm^2mol^{-1}}$\\
Mobility $M_{\phi}$ & $\SI{1.0}{J^{-1}s^{-1}}$ \\
\hline
\end{tabular}
\end{table}

\begin{table}[t]
\caption{Conditions of GAN training}\label{Table2}
\centering
\begin{tabular}{l l}
\hline
Dimensions of latent variables & 2 \\
Distribution of latent variables & Uniform distribution in [0,100] \\
Optimization methods for discriminator & Adam (learning rate, 0.0001) \\ 
Optimization methods for generator & Adam (learning rate, 0.0001) \\ 
Minibatch size & $32$ \\ 
Number of iterations & $1,000,000$ \\
Number of data & $1700$ \\
\hline
\end{tabular}
\end{table}

\begin{figure}[t]%
\centering
\includegraphics[width=0.45\textwidth]{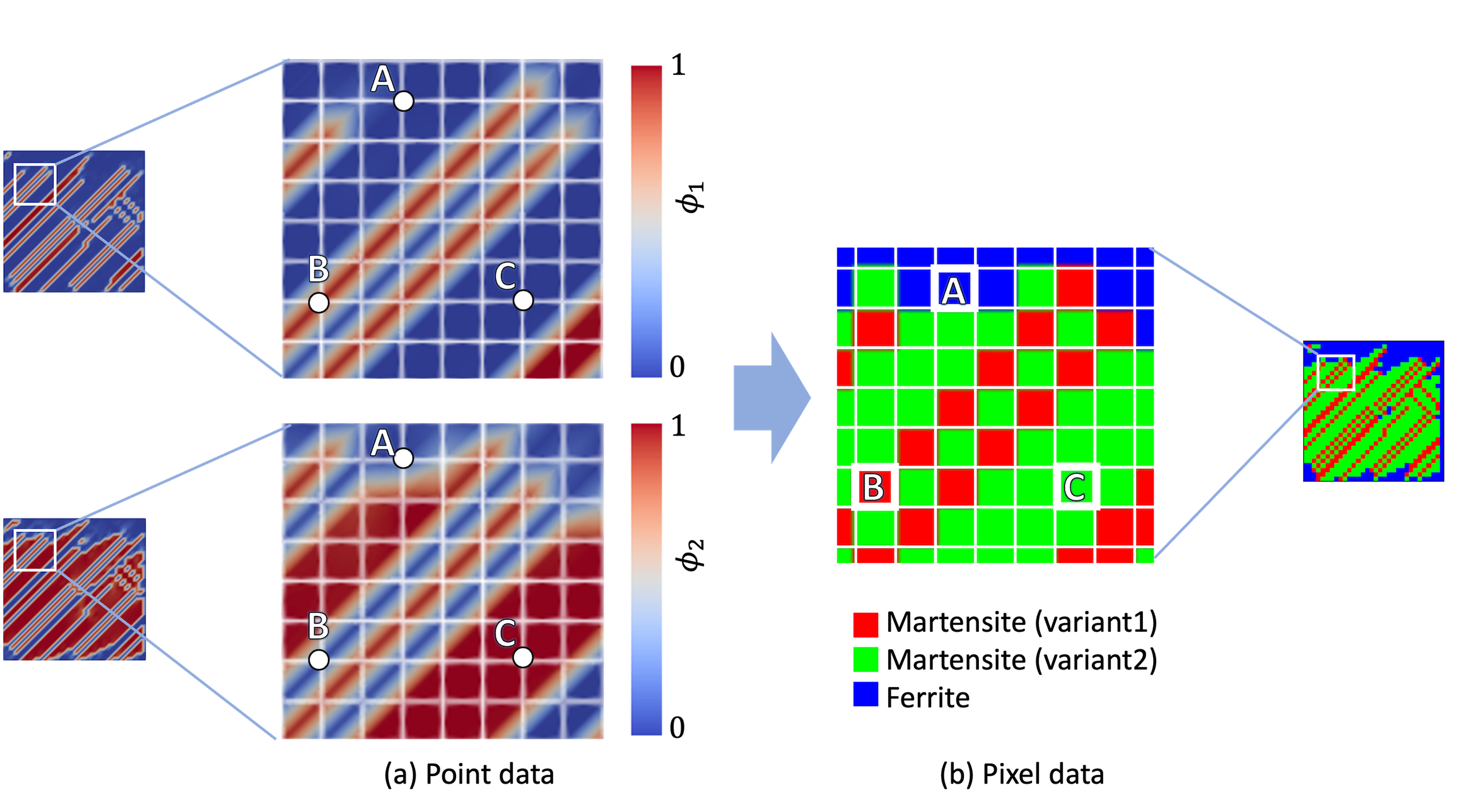}
\caption{Example of converting the results of phase-field analysis to pixel data. (a) Point data. The upper image means the probability that the point is variant$1$ of martensite. The lower image means the probability that the point is variant$2$ of martensite. (b) Pixel data.}
    \label{Fig3}
\end{figure}

\begin{figure}[t]%
\centering
\includegraphics[width=0.6\textwidth]{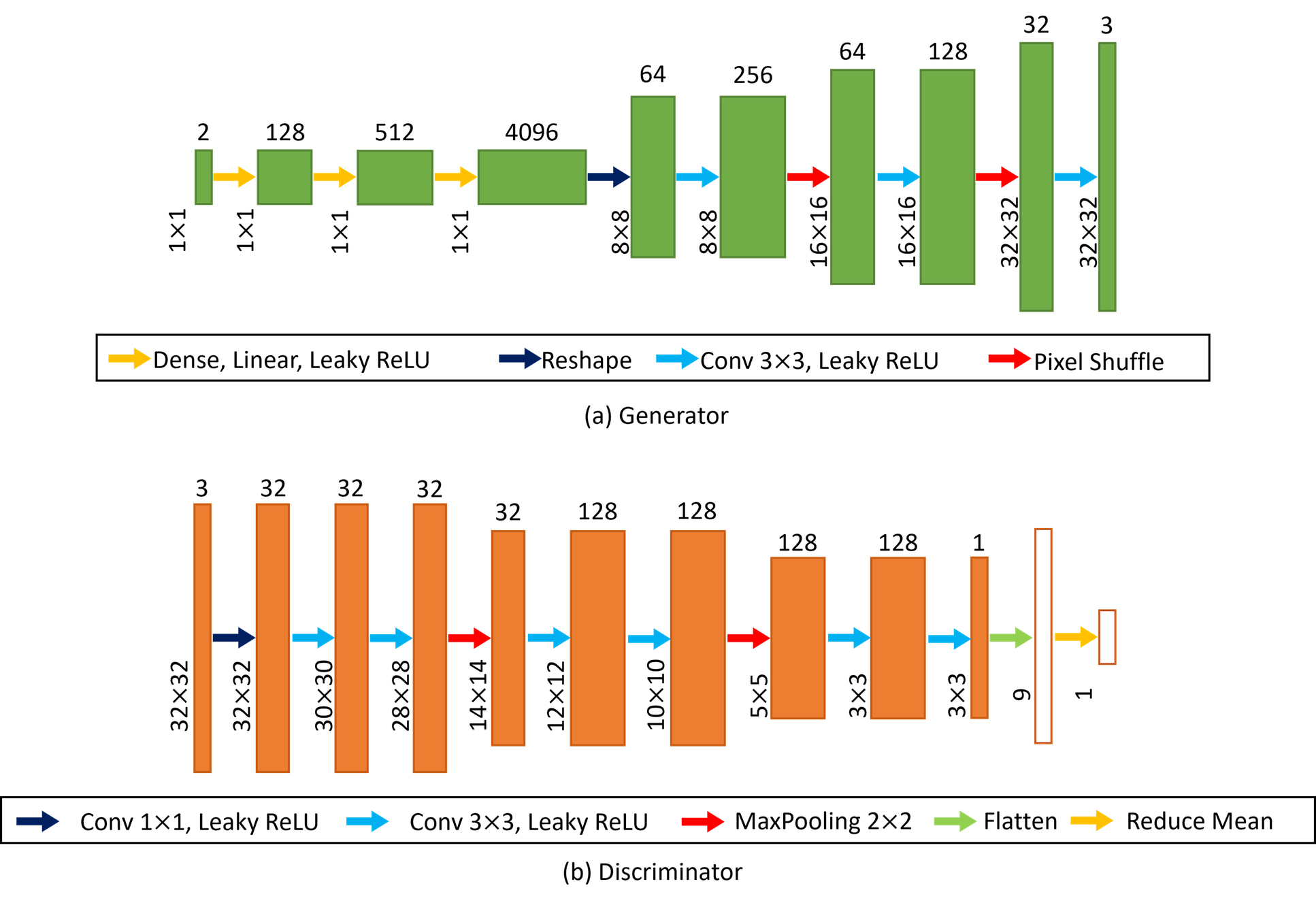}
\caption{Architecture of GAN. (a) Architecture of generator network. (b) Architecture of discriminator network.}
    \label{Fig4}
\end{figure}

\subsection{Generative adversarial network}
\label{subsec2}

GAN \cite{goodfellow2014generative} is a method of adversarial competition and alternating training of two networks, a generator $G$ and a discriminator $D$.
The generator $G$ is a generative model that outputs an image $x^{\prime}$ from a random variable of the latent variable $z$.
On the other hand, the discriminator $D$ outputs a discriminant signal indicating the probability that the image is authentic.
Through repeated training, the generator $G$ is optimized to fool the discriminator $D$ and eventually produces images that are indistinguishable from the real ones.
However, GAN has issues such as learning instability, mode collapse, and the vanishing gradient problem \cite{goodfellow2014distinguishability}.
As a method of solving these problems, Wasserstein GAN (WGAN) \cite{arjovsky2017wasserstein} was developed, which takes the Wasserstein distance approximated by the following equation as the loss function:
\begin{equation}
    W\left(p_r, p_g\right)=\max _{w \in \mathcal{W}} \mathbb{E}_{x \sim p_r}\left[f_w(x)\right]-\mathbb{E}_{z \sim p_z}\left[f_w\left(g_\theta(z)\right)\right],
    \label{eq6}
\end{equation} 
where $W$ is the loss function, $p_r$ is the training image distribution, $p_g$ is the generated image distribution, $p_z$ is the latent variable distribution, $x$ is the image, $z$ is the latent variable, $w$ and $\theta$ are parameters, $\mathcal{W}$ is the parameter space that satisfies Lipschitz continuity, $f_w$ is the function with $w$, and $g_\theta$ is the generated image with $\theta$.
WGAN is used in this study.

As training data for GAN, the results obtained by the phase-field method are converted to pixel data.
Fig. \ref{Fig3} shows an example of converting the results of phase-field analysis to pixel data.
The results of the phase-field analysis are presented as the value of  $\phi_{i}(i=1,2)$ at each grid point, as shown in Fig. \ref{Fig3}a.
On the basis of the value of $\phi_{i}(i=1,2)$ at each grid point, we have to determine whether the pixel is variant$1$ of martensite, variant$2$ of martensite, or a ferrite as shown in Fig. \ref{Fig3}b.
In Fig. \ref{Fig3}b, red pixels indicate variant$1$ of martensite, green pixels indicate variant$2$ of martensite, and blue pixels indicate a ferrite.

As can be seen from Fig. \ref{Fig3}a, $\phi_{i}(i=1,2)$ takes values from $0$ to $1$.
Both $\phi_{1}$ and $\phi_{2}$ can take non zero values, and one of these is sufficiently smaller than the other.
On the basis of these features, point data are converted to pixel data according to the following criteria by using the variable $P$, which is $P=1$ if a pixel is a variant$1$ of martensite, $P=2$ if a pixel is a variant$2$ of martensite, and $P=0$ if a pixel is ferrite.
\begin{equation}
  P =
  \begin{cases}
    1 & \text{if $\phi_{1}=\mathrm{max}(\phi_{1},\phi_{2},\phi_{0})$}, \\
    2 & \text{if $\phi_{2}=\mathrm{max}(\phi_{1},\phi_{2},\phi_{0})$}, \\
    0 & \text{if $\phi_{0}=\mathrm{max}(\phi_{1},\phi_{2},\phi_{0})$},
  \end{cases}
  \label{eq7}
\end{equation}
where $\phi_{0}=1-\phi_{1}-\phi_{2}$.
At point A in Fig. \ref{Fig3}a, $\phi_{0}=\mathrm{max}(\phi_{1},\phi_{2},\phi_{0})$ is satisfied.
From Eq. \eqref{eq7}, point A corresponds to ferrite ($P=0$), and pixel A in Fig. \ref{Fig3}b is blue.
Similarly, point B in Fig. \ref{Fig3}a corresponds to variant$1$ of martensite ($P=1$), and pixel B in Fig. \ref{Fig3}b becomes red.
Point C in Fig. \ref{Fig3}a corresponds to variant$2$ of martensite ($P=2$), and pixel B in Fig. \ref{Fig3}b becomes green.
Thus, the converted pixel data have three channels, RGB, and the size of the training data is set to $\SI{32}{pixels}\times\SI{32}{pixels}\times\SI{3}{channels}$.
The architecture of GAN is shown in Fig. \ref{Fig4} and the training conditions of GAN are shown in Table \ref{Table2}.

\begin{figure}[t]%
\centering
\includegraphics[width=0.8\textwidth]{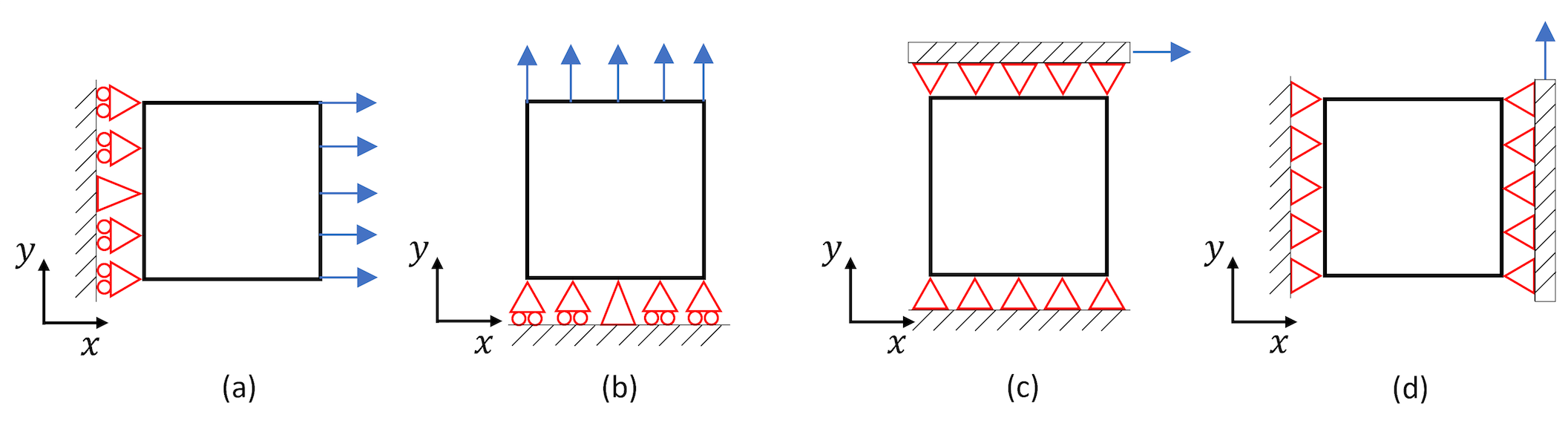}
\caption{Boundary conditions for FEM. (a) Tensile toward $x$ direction. (b) Tensile toward $y$ direction. (c) Shear toward $x$ direction. (d) Shear toward $y$ direction}
\label{Fig5}
\end{figure}

\section{Construction of CNN}\label{sec4}
In the inverse analysis framework shown in Fig. \ref{Fig1}, CNN is used to predict the maximum stress $\sigma_{\mathrm{max}}$ and the working limit strain $\varepsilon_{\mathrm{lim}}$ from DP steel microstructures.
The training data for the input of CNN are parts of the DP steel microstructures generated in Section \ref{subsec1}.
The training data for the output of CNN are $\sigma_{\mathrm{max}}$ and $\varepsilon_{\mathrm{lim}}$ calculated by the dislocation-crystal plasticity FEM.

The crystal plasticity FEM is suitable for the evaluation of the mechanical properties of metals because it takes into consideration crystal information such as grain size and crystal orientation.
The problem with the crystal plasticity FEM is, however, its high computational cost \cite{yamanaka2020deep,knezevic2009crystal}.
To enable iterative random searches that repeat the processes of generating structures and evaluating their mechanical properties, as the one shown in Fig. \ref{Fig1}, the dislocation-crystal plasticity FEM should be replaced by CNN.

In section \ref{subsec3}, we describe the preparation of $\sigma_{\mathrm{max}}$ and $\varepsilon_{\mathrm{lim}}$ using the dislocation-crystal plasticity FEM, and in Section \ref{subsec4}, we describe the training of CNN.

\subsection{Dislocation-crystal plasticity finite element method}\label{subsec3}
Dislocation-crystal plasticity FEM is used to calculate the maximum stress $\sigma_{\mathrm{max}}$ and the working limit strain $\varepsilon_{\mathrm{lim}}$, which are obtained by deforming a ductile metallic specimen at a constant deformation rate.
The maximum stress $\sigma_{\mathrm{max}}$ is the maximum nominal stress that can be applied to the specimen and represents strength.
The working limit strain $\varepsilon_{\mathrm{lim}}$ is the true strain at the start of necking and represents ductility.
The working limit strain $\varepsilon_{\mathrm{lim}}$ is the true strain that satisfies the following equation expressed by the true stress $\bar{\sigma}$ and the true strain ${\bar{\varepsilon}}$:
\begin{equation}
\frac{d\bar{\sigma}}{d\bar{\varepsilon}}=\bar{\sigma}.
\label{eq8}
\end{equation}

The dislocation-crystal plasticity model \cite{asaro1977strain,asaro1979geometrical} is used as the constitutive law of FEM.
Specifically, it is expressed as:
\begin{equation}
\overset{\circ}{\boldsymbol{\sigma}}=\boldsymbol{C}:\boldsymbol{D}-\sum_{\alpha=1}^{N} \dot{\gamma}^{(\alpha)}\left[\boldsymbol{C}:\boldsymbol{p}^{(\alpha)}+\boldsymbol{w}^{(\alpha)} \boldsymbol{\sigma}-\boldsymbol{\sigma} \boldsymbol{w}^{(\alpha)}\right],
\label{eq9}
\end{equation}
where $\overset{\circ}{\boldsymbol{\sigma}}$ is the Jaumann velocity of stress, $\boldsymbol{C}$ is the elastic modulus, $\boldsymbol{D}$ is the deformation rate, $\dot{\gamma}^{(\alpha)}$ is the slip rate of slip system $\alpha$, and $\boldsymbol{p}^{(\alpha)}$ and $\boldsymbol{w}^{(\alpha)}$ are the tensors determined when a slip system is determined. 
The hardening law for the shear slip rate $\dot{\gamma}^{(\alpha)}$ is the exponential law used by Hutchinson \cite{hutchinson1976bounds} and Pan-Rice \cite{pan1983rate} as follows:
\begin{equation}
    \dot{\gamma}^{(\alpha)}=\dot{\gamma}^{(\alpha)}_{0}\mathrm{sgn}(\tau^{(\alpha)})\left\lvert\frac{\tau^{(\alpha)}}{g^{(\alpha)}}\right\rvert^{\frac{1}{m}},
    \label{eq10}
\end{equation}
where $\mathrm{sgn}$ is the sign function, $\dot{\gamma}^{(\alpha)}_{0}$ is the reference slip rate, $\tau^{(\alpha)}$ is the resolved shear stress, $g^{(\alpha)}$ is the flow stress, and $m$ is the strain-rate sensitivity.
In crystal plasticity theory, the evolution equation of the flow stress is expressed as:
\begin{equation}
    \dot{g}^{(\alpha)} = \sum_{\beta=1}^{N} h^{(\alpha\beta)} \left\lvert\dot{\gamma}^{(\beta)}\right\rvert,
    \label{eq11}
\end{equation}
where $h^{(\alpha\beta)}$ is the dislocation-dependent hardening modulus.

In order to obtain the relationship between the hardening modulus and the dislocation density, we introduce the density of the geometrically necessary (GN) dislocations $\rho_{\mathrm{G}}$ and the density of the statistically stored (SS) dislocations $\rho_{\mathrm{S}}$ \cite{kujirai2020modelling,ohashi1994numerical,kimura2020crystal}.
Here, the definitions of the screw and edge components of the GN dislocation density are expressed as:
\begin{equation}
    \dot{\rho}_{\mathrm{G,screw}}^{(\alpha)} =\frac{1}{\tilde{b}}\nabla\dot{\gamma}^{(\alpha)}\cdot\boldsymbol{t}^{(\alpha)},
    \label{eq12}
\end{equation}
\begin{equation}
    \dot{\rho}_{\mathrm{G,edge}}^{(\alpha)} =-\frac{1}{\tilde{b}}\nabla\dot{\gamma}^{(\alpha)}\cdot\boldsymbol{s}^{(\alpha)},
    \label{eq13}
\end{equation}
where $\tilde{b}$ is the magnitude of Burgers vector, $\boldsymbol{s}^{(\alpha)}$ is the unit vector in the slip direction, and $\boldsymbol{t}^{(\alpha)}$ is the unit binormal vector defined by $\boldsymbol{t}^{(\alpha)}=\boldsymbol{s}^{(\alpha)}\times\boldsymbol{m}^{(\alpha)}$, and $\boldsymbol{m}^{(\alpha)}$ is the unit vector normal to the slip plane.
The evolution equation of the SS dislocation density are expressed as:
\begin{equation}
    \dot{\rho}_{\mathrm{S}}^{(\alpha)} =\frac{c}{\tilde{b}L^{(\alpha)}}\left\lvert\dot{\gamma}^{(\alpha)}\right\rvert,
    \label{eq14}
\end{equation}
where $c$ is a numerical parameter on the order of $1$ and $L^{(\alpha)}$ is the dislocation mean free path.

The relationship between the flow stress and the dislocation density can be written as \cite{bailey1960dislocation}:
\begin{equation}
    g^{(\alpha)}= \tau_{y}^{(\alpha)}+a\mu\tilde{b}\sum_{\beta}\varOmega^{(\alpha\beta)}\sqrt{\rho^{(\beta)}_{\mathrm{S}}},
    \label{eq15}
\end{equation}
where $\tau_{y}^{(\alpha)}$ is the reference shear stress, $a$ is a numerical parameter on the order of $0.1$, $\mu$ is shearing modulus, and $\varOmega^{(\alpha\beta)}$ is the matrix representing the dislocation interaction between slip systems $\alpha$ and $\beta$.
Comparing Eq. \eqref{eq11} and the time derivative of Eq. \eqref{eq15}, we obtain the relationship as follows:
\begin{equation}
    h^{(\alpha\beta)}= \frac{a\mu\tilde{b}\varOmega^{(\alpha\beta)}c}{2\tilde{b}L^{(\beta)}\sqrt{\rho^{(\beta)}_{\mathrm{S}}}}.
    \label{eq16}
\end{equation}
The dislocation mean free path $L^{(\beta)}$ is expressed as:
\begin{equation}
    L^{(\beta)}= \frac{c^{*(\beta)}}{\sqrt{\sum_{\gamma} \omega^{(\beta\gamma)}\left(\rho^{(\gamma)}_{\mathrm{G}}+\rho^{(\gamma)}_{\mathrm{S}}\right)}},
    \label{eq17}
\end{equation}
where $c^{*(\beta)}$ is the dislocation mobility and $\omega^{(\beta\gamma)}$ is the dislocation interaction matrix excluding the effect of self-hardening.

For the FEM, we determined whether each element is a martensite with variant$1$, a martensite with variant$2$, or a ferrite based on the basis of the values of $\phi_{i}(i=1,2)$ obtained by phase-field analysis.
The simulation conditions for the FEM are shown in Table \ref{Table3} and the material information in Table \ref{Table4}.
Anisotropy is taken into account and the analysis is performed under four boundary conditions, namely, tensile toward the $x$ direction, tensile toward the $y$ direction, shear toward the $x$ direction, and shear toward the $y$ direction, as shown in Fig. \ref{Fig5}.

\begin{table}[t]
\caption{Calculation conditions for FEM}\label{Table3}
\centering
\begin{tabular}{l l}
\hline
Size & $\SI{31}{\ \um}\times\SI{31}{\ \um}$\\
Number of grid points & $32\times32$\\
Strain rate & $1.0\times10^{-4}\si{\, s^{-1}}$\\
Crystal orientation & [0,0,10] \\
2D analysis conditions & Plane strain \\
\hline
\end{tabular}
\end{table}

\begin{table}[t]
\caption{Martensite and ferrite material information used in FEM}\label{Table4}
\centering
\begin{tabular}{l l l}
\hline
& Martensite &  Ferrite \\\hline
Young's modulus & $\SI{237.3}{\ \GPa}$ & $\SI{205.9}{\ \GPa}$\\
Poisson's ratio & $0.333$ & $0.3$ \\
Initial dislocation density & $\SI{1.0E3}{\ \um^{-2}}$ & $\SI{1.0}{\ \um^{-2}}$ \\
Strain rate sensitivity & $0.007$ & $0.01$ \\ 
Reference strain rate & $1.0\,\mathrm{ms^{-1}}$ &$1.0\,\mathrm{ms^{-1}}$\\
\hline
\end{tabular}
\end{table}

\subsection{Convolutional neural network}\label{subsec4}
CNN is commonly used for image recognition because it can acquire important features for prediction by condensing spatial information such as images \cite{bishop2006pattern,goodfellow2016deep}.
Images have three-dimensional information (vertical, horizontal, and channel), but the fully connected layer requires that the information of an image be converted to one dimension at input, making it impossible to effectively use the original spatial information.
In contrast, CNN does not lose spatial information of the image as follows:
\begin{equation}
    y_{ijk}=\sum_{l}\sum_{p}\sum_{q}W_{pqkl}x_{(i+p)(j+q)l}+b_{k},
    \label{eq18}
\end{equation}
where $\boldsymbol{W}$ denotes the kernel and $\boldsymbol{b}$ the bias term.

\begin{figure}[t]%
\centering
\includegraphics[width=0.6\textwidth]{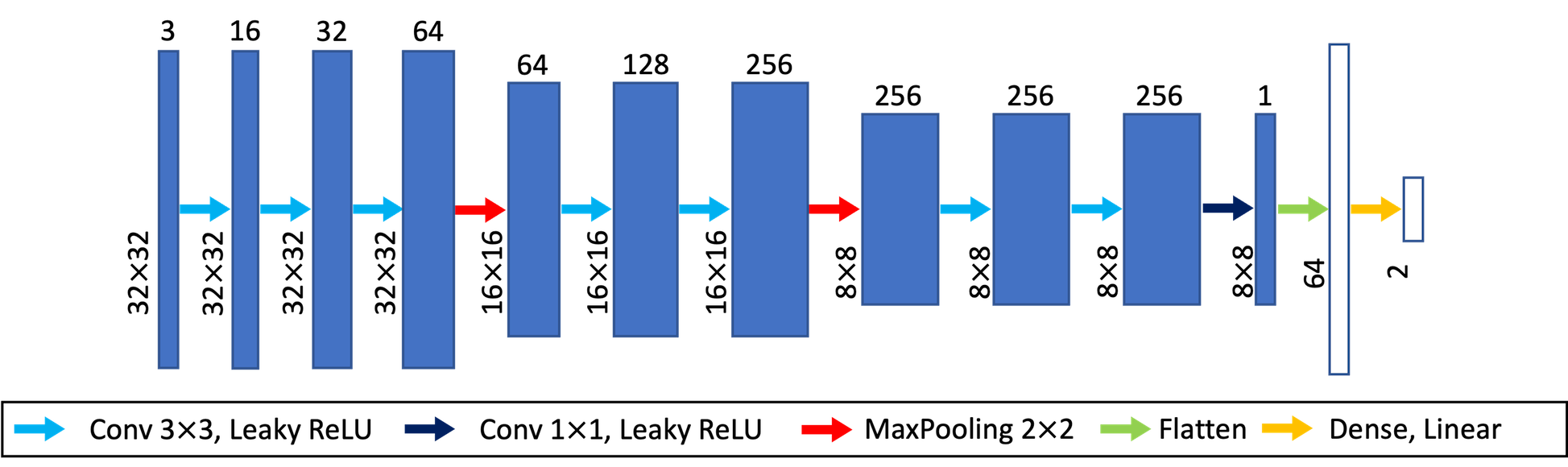}
\caption{Architecture of CNN.}
\label{Fig6}
\end{figure}

Here, we construct CNN that outputs the maximum stress $\sigma_{\mathrm{max}}$ and the working limit strain $\varepsilon_{\mathrm{lim}}$ when DP steel microstructures are given.
The architecture of CNN is shown in Fig. \ref{Fig6} and the training conditions of CNN are shown in Table \ref{Table5}.

\begin{table}[t]%
\caption{CNN training conditions}\label{Table5}
\centering
\begin{tabular}{l l}
\hline
Loss function & Mean Squared Error (MSE) \\ 
Optimization method & Adam (learning rate, 0.0001) \\ 
Minibatch size  & 1 (tensile deformation) \\ 
 & 4 (shear deformation) \\ 
Number of iterations & 1200 \\
Number of training data & 96\\
Number of validation data & 10\\
Number of test data & 10\\
\hline
\end{tabular}
\end{table}

\section{Results and Discussion}\label{sec5}

\subsection{Output of DP steel microstructures based on latent variables}\label{subsec5}
\begin{figure}[t]%
\centering
\includegraphics[width=0.45\textwidth]{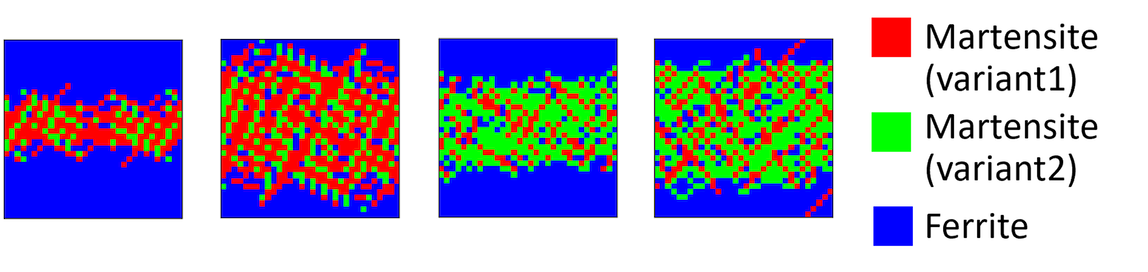}
\caption{Examples of DP steel microstructures generated by phase-field method.}
\label{Fig7}
\end{figure}

The phase-field method is used to generate the DP steel microstructures as shown in Fig. \ref{Fig7}.
GAN is constructed using the generated microstructures as training data.
The shape of training data of GAN is $\SI{32}{pixels}\times\SI{32}{pixels}\times\SI{3}{channels}$.
Each channel corresponds to ferrite, variant$1$ of martensite, and variant$2$ of martensite, respectively.
The number of training data is $1700$.
The dimension of the latent variable is set to $2$ and the distribution of the latent variable is set to a uniform distribution in the range of $[0, 100]$.
The number of training iterations is set to $1,000,000$.
The generator $G$ is trained once every $10$ iterations and the discriminator $D$ is trained $9$ times every $10$ iterations.
The size of the mini batch is $32$, the optimization method is Adam, and the learning rate is $0.0001$.

\begin{figure}[t]%
\centering
\includegraphics[width=0.4\textwidth]{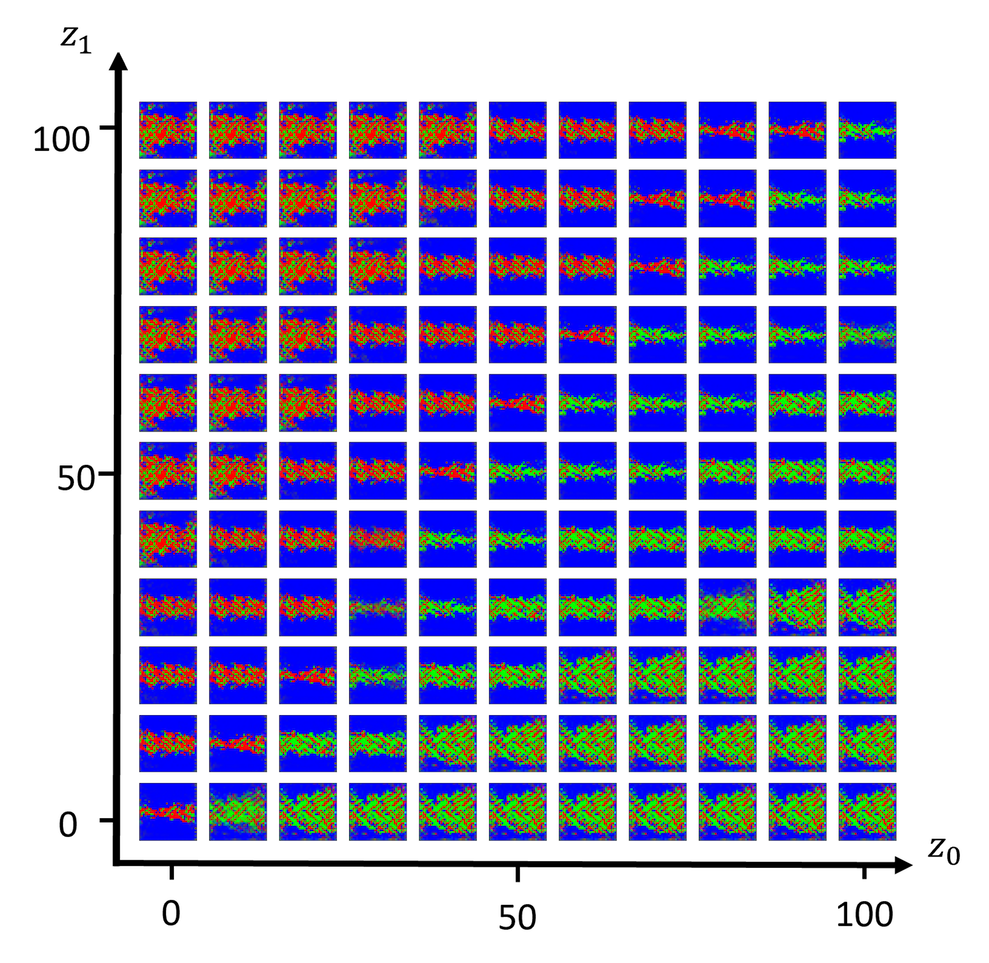}
\caption{Results of GAN. The features of the training data are extracted and obtained in the latent variable space.}
\label{Fig8}
\end{figure}

Changes in the DP steel microstructures in the two-dimensional latent variable space of the trained GAN are shown in Fig. \ref{Fig8}.
In Fig. \ref{Fig8}, the fraction of martensite decreases at around $z_{1}=z_{0}$, and from the line $z_{1}=z_{0}$, the fraction of variant$2$ increases as $z_{0}$ increases and $z_{1}$ decreases.
Conversely, the smaller $z_{0}$ and the larger $z_{1}$ are, the larger the fraction of variant$1$ is.
Thus, the microstructures vary with the values of the latent variable $\boldsymbol{z}$.
We also find that similar microstructures are distributed closely together in the latent variable space.

\begin{figure}[t]%
\centering
\includegraphics[width=0.47\textwidth]{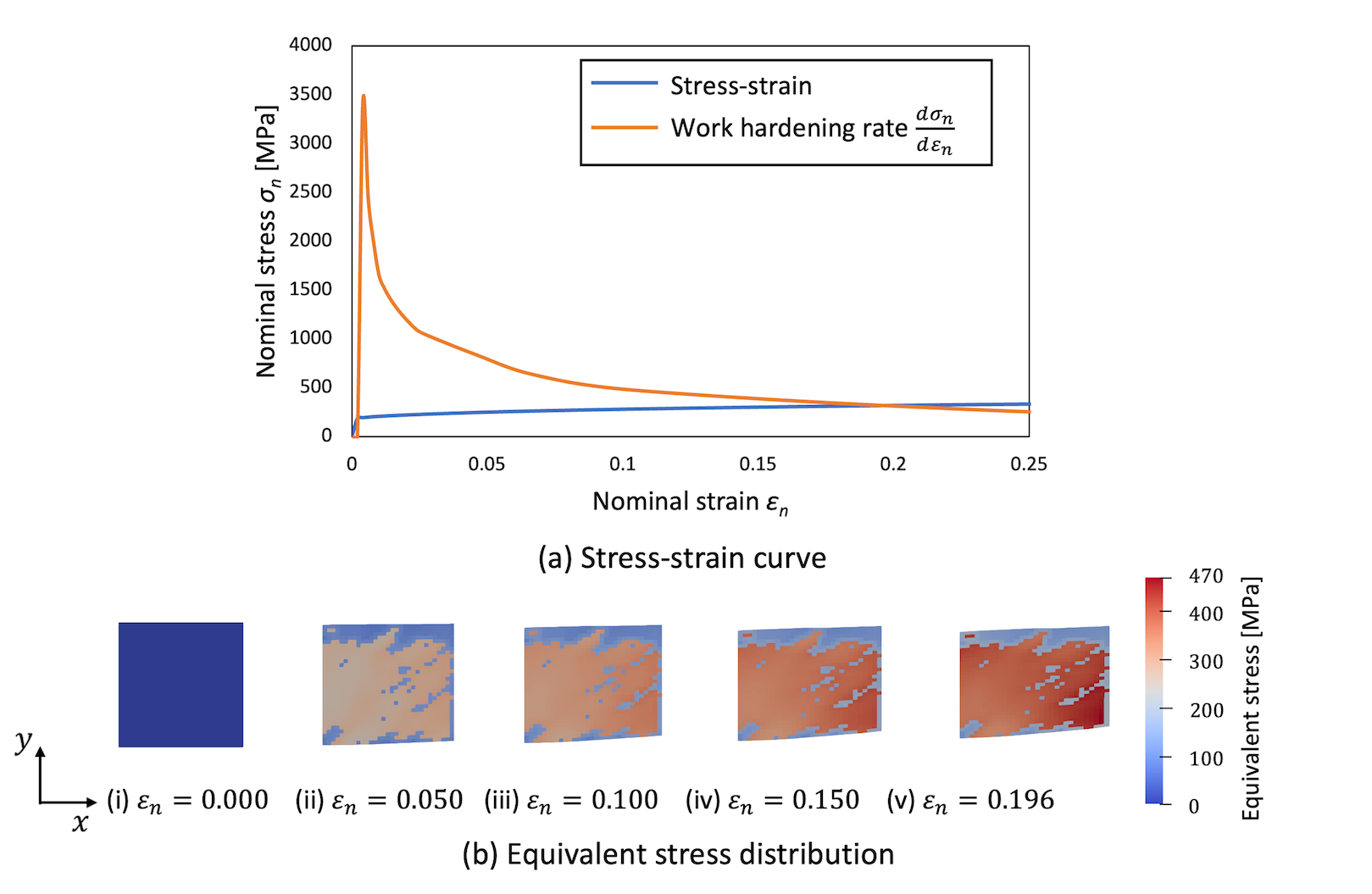}
\caption{Example of FEM results in the case of tensile toward the $x$ direction. (a) The upper figure shows the stress--strain and work hardening curves. (b) The lower figure shows the historical distributions of the equivalent stress.}
\label{Fig9}
\end{figure}

\begin{figure}[t]%
\centering
\includegraphics[width=0.47\textwidth]{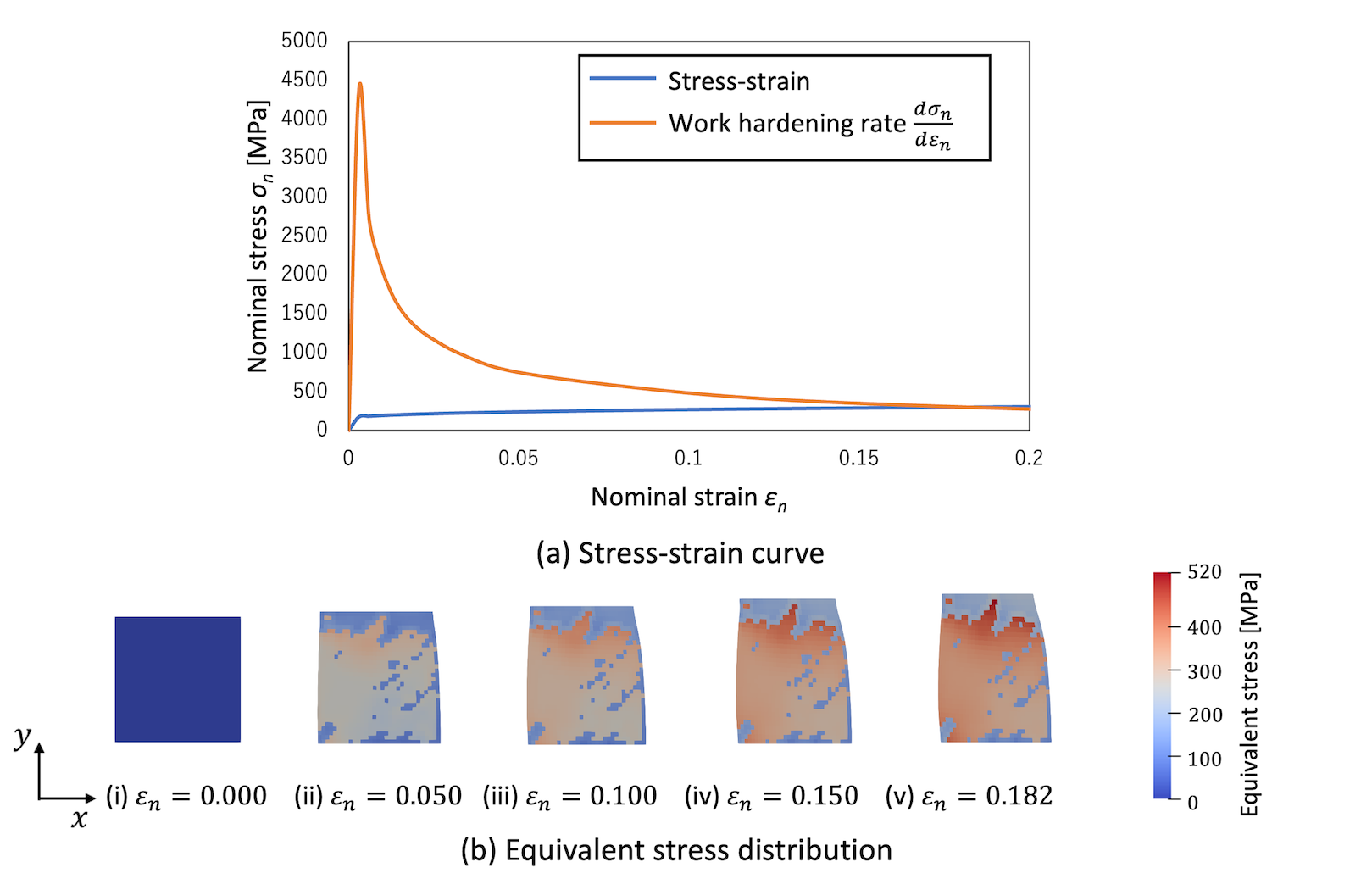}
\caption{Example of FEM results in the case of tensile toward the $y$ direction. (a) The upper figure shows the stress--strain and work hardening curves. (b) The lower figure shows the historical distributions of the equivalent stress.}
\label{Fig10}
\end{figure}

\begin{figure}[t]%
\centering
\includegraphics[width=0.47\textwidth]{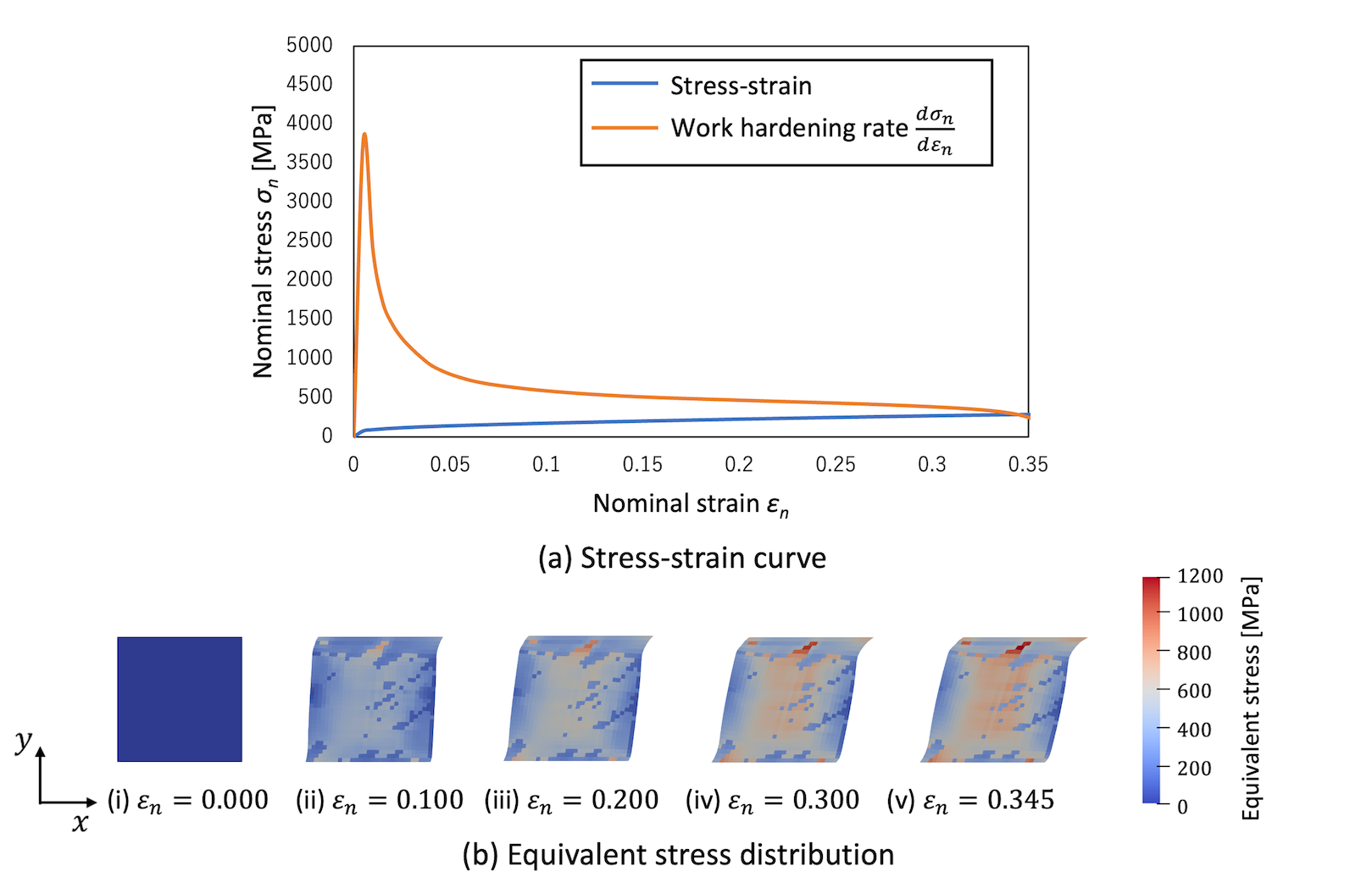}
\caption{Example of FEM results in the case of shear toward the $x$ direction. (a) The upper figure shows the stress--strain and work hardening curves. (b) The lower figure shows the historical distributions of the equivalent stress.}
\label{Fig11}
\end{figure}

\begin{figure}[t]%
\centering
\includegraphics[width=0.47\textwidth]{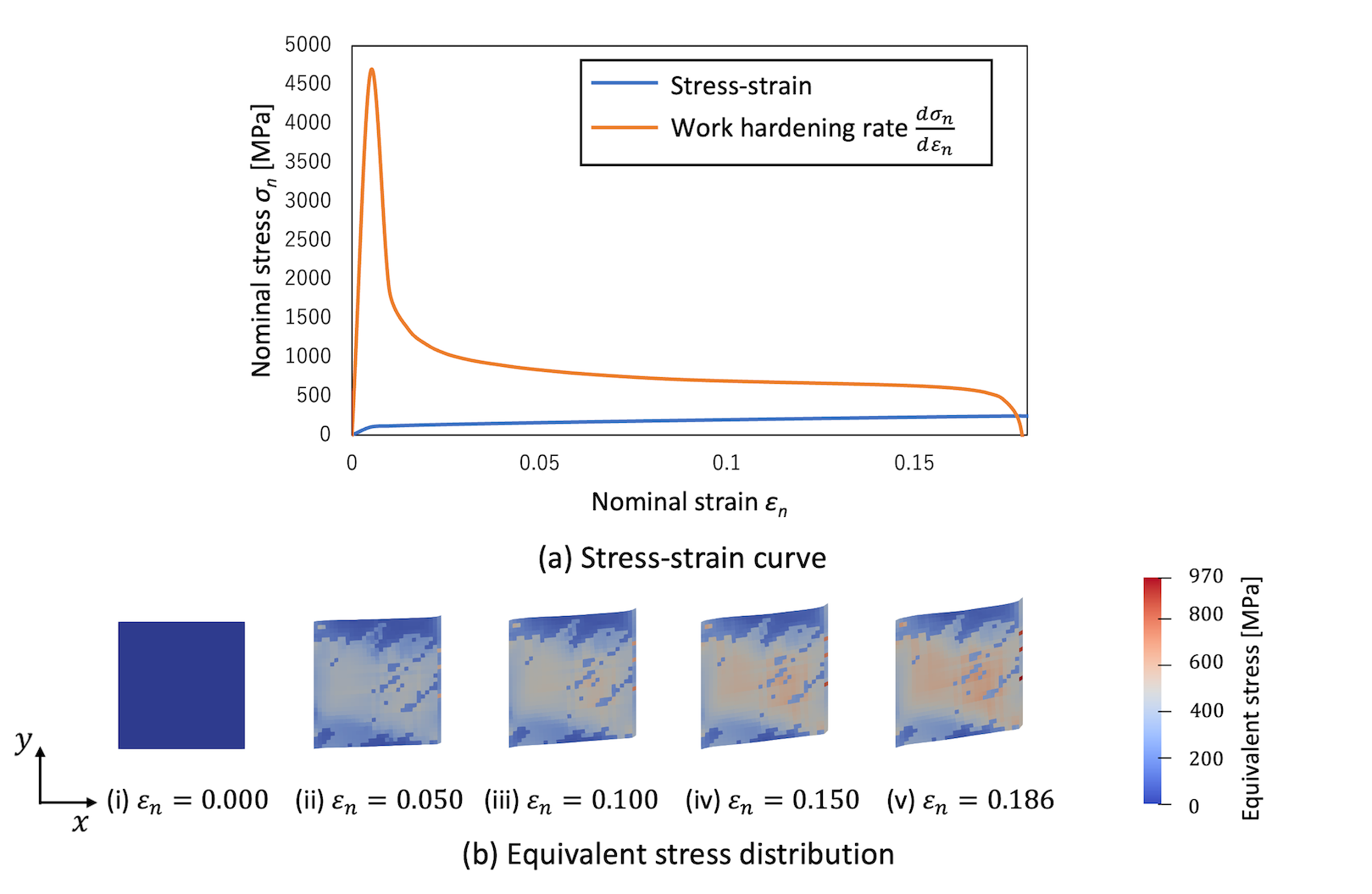}
\caption{Example of FEM results in the case of shear toward the $y$ direction. (a) The upper figure shows the stress--strain and work hardening curves. (b) The lower figure shows the historical distributions of the equivalent stress.}
\label{Fig12}
\end{figure}

\subsection{Output of maximum stress and working limit strain based on microstructures}\label{subsec6}
The FEM is performed on the DP steel microstructures prepared in Section \ref{subsec5} to obtain the maximum stress $\sigma_{\mathrm{max}}$ and the working limit strain $\varepsilon_{\mathrm{lim}}$.
Here, an example of the results of FEM in the case of tensile toward the $x$ direction is shown in Fig. \ref{Fig9}.
Fig. \ref{Fig9} shows the stress--strain curve and the equivalent stress distribution during deformation.
From the stress--strain curve, a work hardening curve is derived. 
From the intersection of the stress--strain and work hardening curves, the maximum stress $\sigma_{\mathrm{max}}$ and the working limit strain $\varepsilon_{\mathrm{lim}}$ are obtained.
The equivalent stress distribution indicates that the martensitic phase carried a higher stress than the ferritic phase.
For the same microstructure as in the case of tensile toward the $x$ direction, the stress--strain curves and equivalent stress distributions in the case of tensile toward the $y$ direction, shear toward the $x$ direction, and shear toward the $y$ direction are shown in Figs. \ref{Fig10}, \ref{Fig11}, and \ref{Fig12}, respectively.
Both $\sigma_{\mathrm{max}}$ and $\varepsilon_{\mathrm{lim}}$ are obtained similarly to the case of tensile toward the $x$ direction.
The equivalent stress distribution indicates the same trend as that in the case of tensile toward the $x$ direction, that is, the martensitic phase carried a higher stress than the ferritic phase.

\begin{figure}[t]%
\centering
\includegraphics[width=0.35\textwidth]{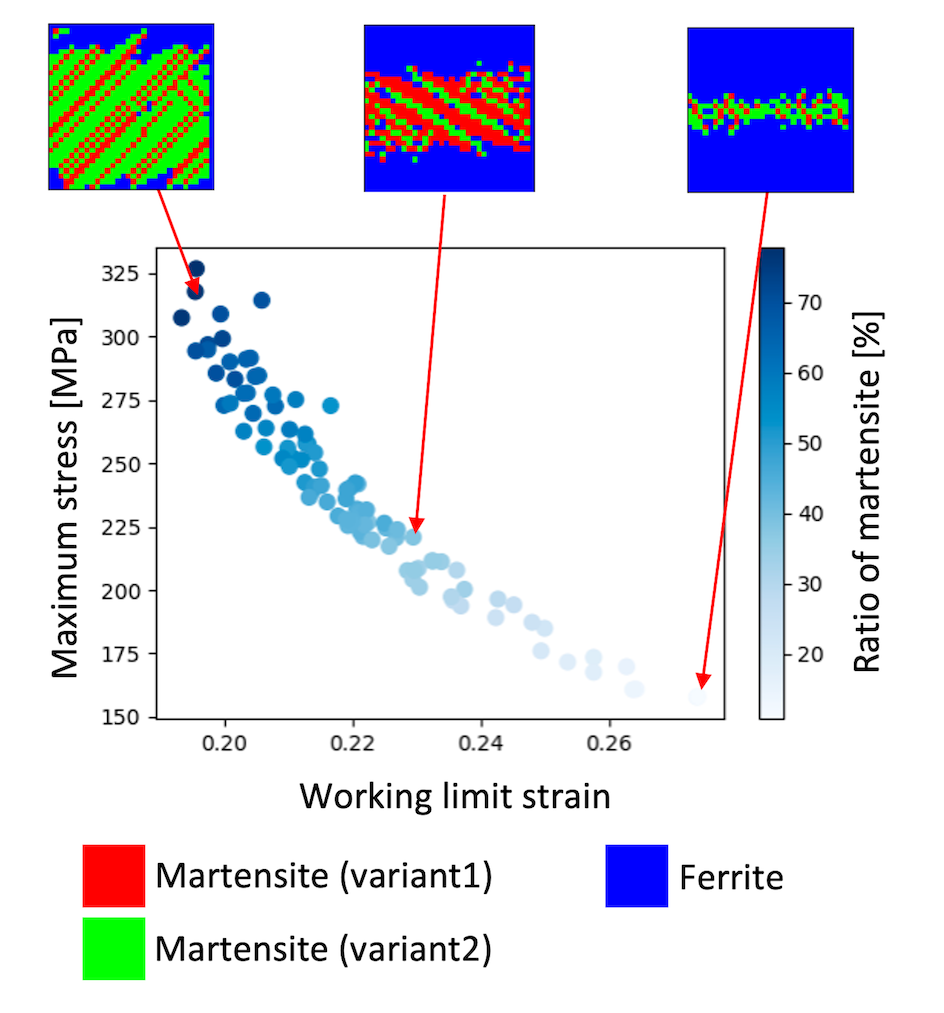}
\caption{Relationship between maximum stress and work limit strain for tensile toward $x$ direction.}
\label{Fig13}
\end{figure}

\begin{figure}[t]%
\centering
\includegraphics[width=0.35\textwidth]{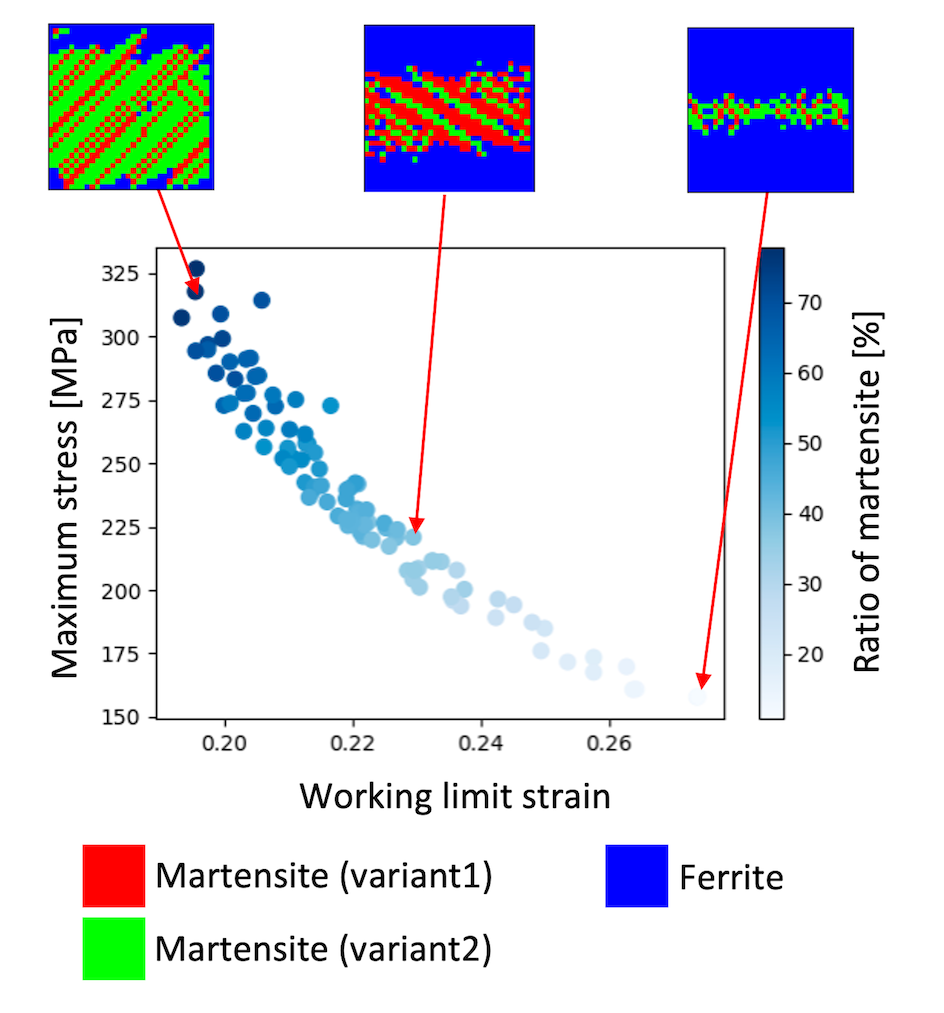}
\caption{Relationship between maximum stress and work limit strain for tensile toward $y$ direction.}
\label{Fig14}
\end{figure}

\begin{figure}[t]%
\centering
\includegraphics[width=0.35\textwidth]{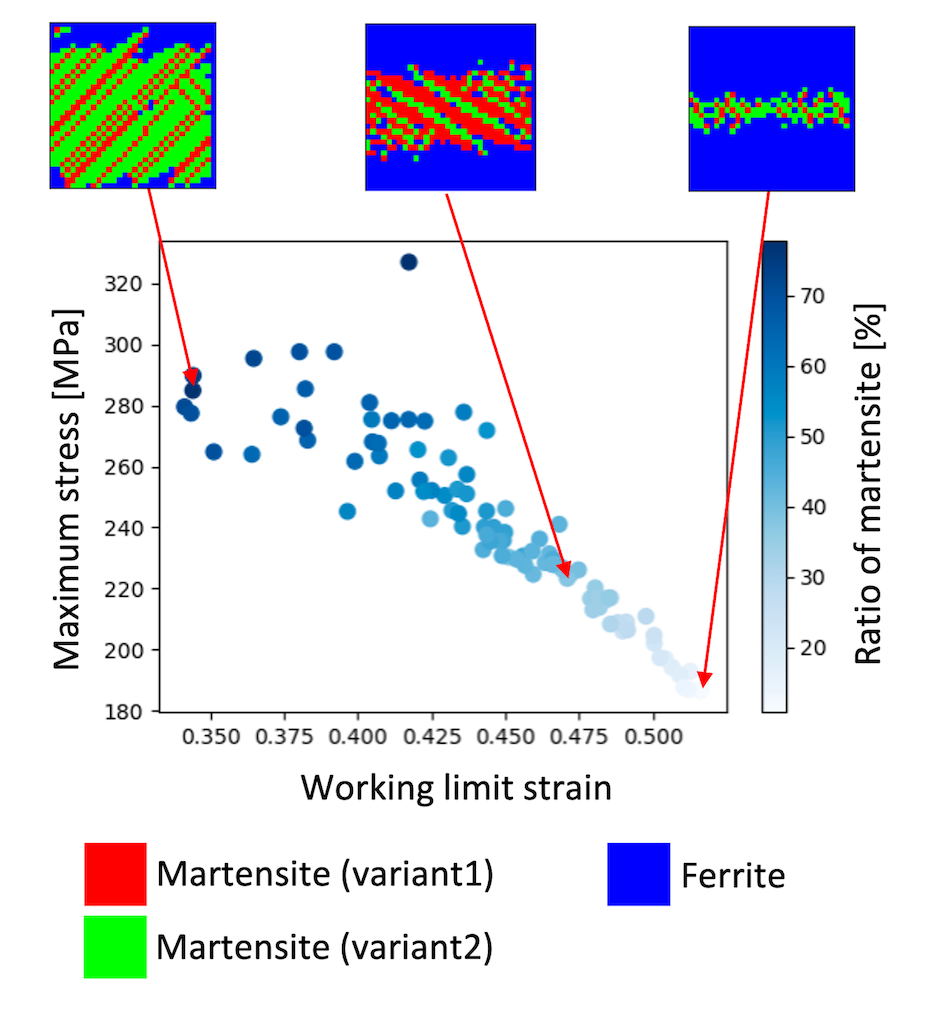}
\caption{Relationship between maximum stress and work limit strain for shear toward $x$ direction.}
\label{Fig15}
\end{figure}

\begin{figure}[t]%
\centering
\includegraphics[width=0.35\textwidth]{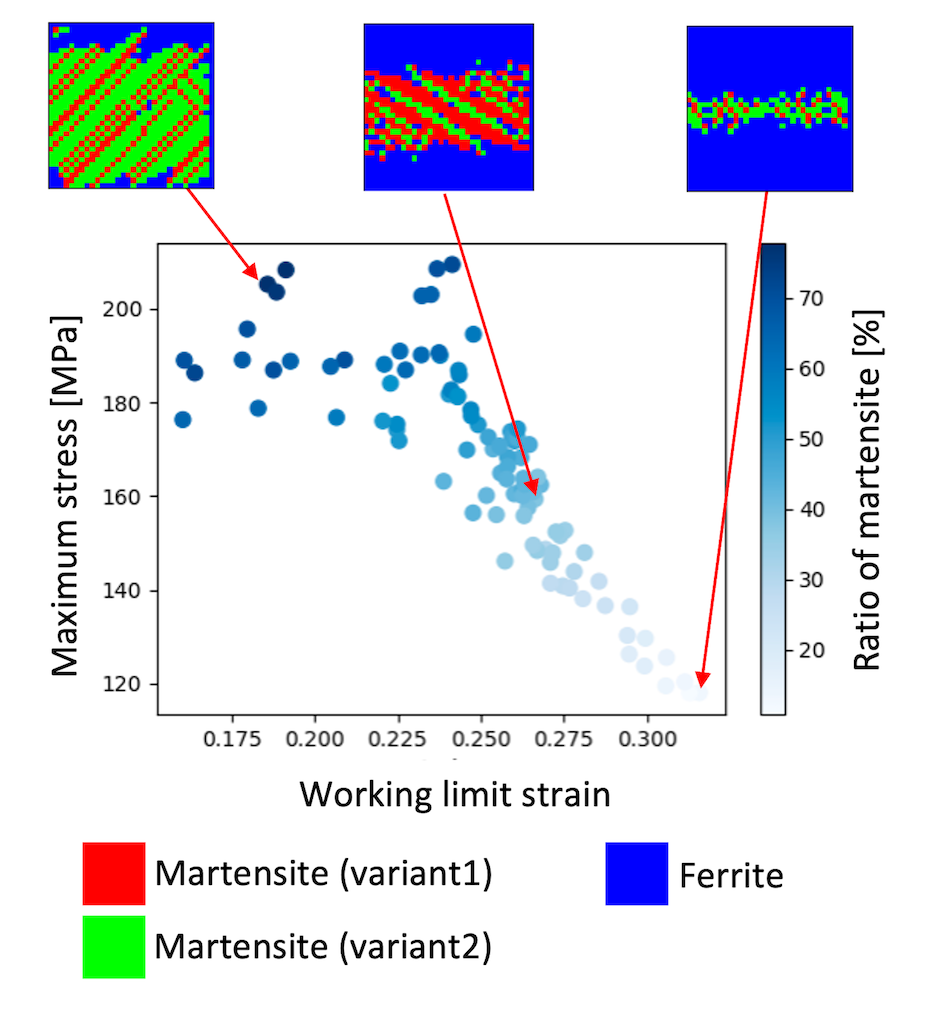}
\caption{Relationship between maximum stress and work limit strain for shear toward $y$ direction.}
\label{Fig16}
\end{figure}

Fig. \ref{Fig13} shows the relationship between $\sigma_{\mathrm{max}}$ and $\varepsilon_{\mathrm{lim}}$ for all training data in the case of tensile toward the $x$ direction.
It can be seen from Fig. \ref{Fig13} that the trade-off between strength $\sigma_{\mathrm{max}}$ and ductility $\varepsilon_{\mathrm{lim}}$ is also observed in the FEM results.
Similarly, the relationships between $\sigma_{\mathrm{max}}$ and $\varepsilon_{\mathrm{lim}}$ for all training data in the case of tensile toward the $y$ direction, the shear toward the $x$ direction, and the shear toward the $y$ direction are shown in Figs. \ref{Fig14}, \ref{Fig15}, and \ref{Fig16}, respectively.
Although there is a larger variation than in the case of tensile toward the $x$ direction, the trade-off relationship follows the same trend as that in the case of tensile toward the $x$ direction.

Using these results of FEM, we construct CNN.
CNN is trained separately for each deformation mode.
Therefore, four CNNs are trained.
The input of CNN is the DP steel microstructures obtained by the phase-field method.
The output of CNN is the maximum stress $\sigma_{\mathrm{max}}$ and the working limit strain $\varepsilon_{\mathrm{lim}}$ obtained by FEM.
The number of training data is $96$ for each deformation mode.
The loss function is the mean squared error, the optimization method is Adam, and the learning rate is $0.0001$.
The size of mini batch is set to $1$ in the case of tensile toward the $x$ direction and tensile toward the $y$ direction, and $4$ in the case of shear toward the $x$ direction and shear toward the $y$ direction.
The number of iterations is set to $1200$.

\begin{figure}[ht]%
\centering
\includegraphics[width=0.45\textwidth]{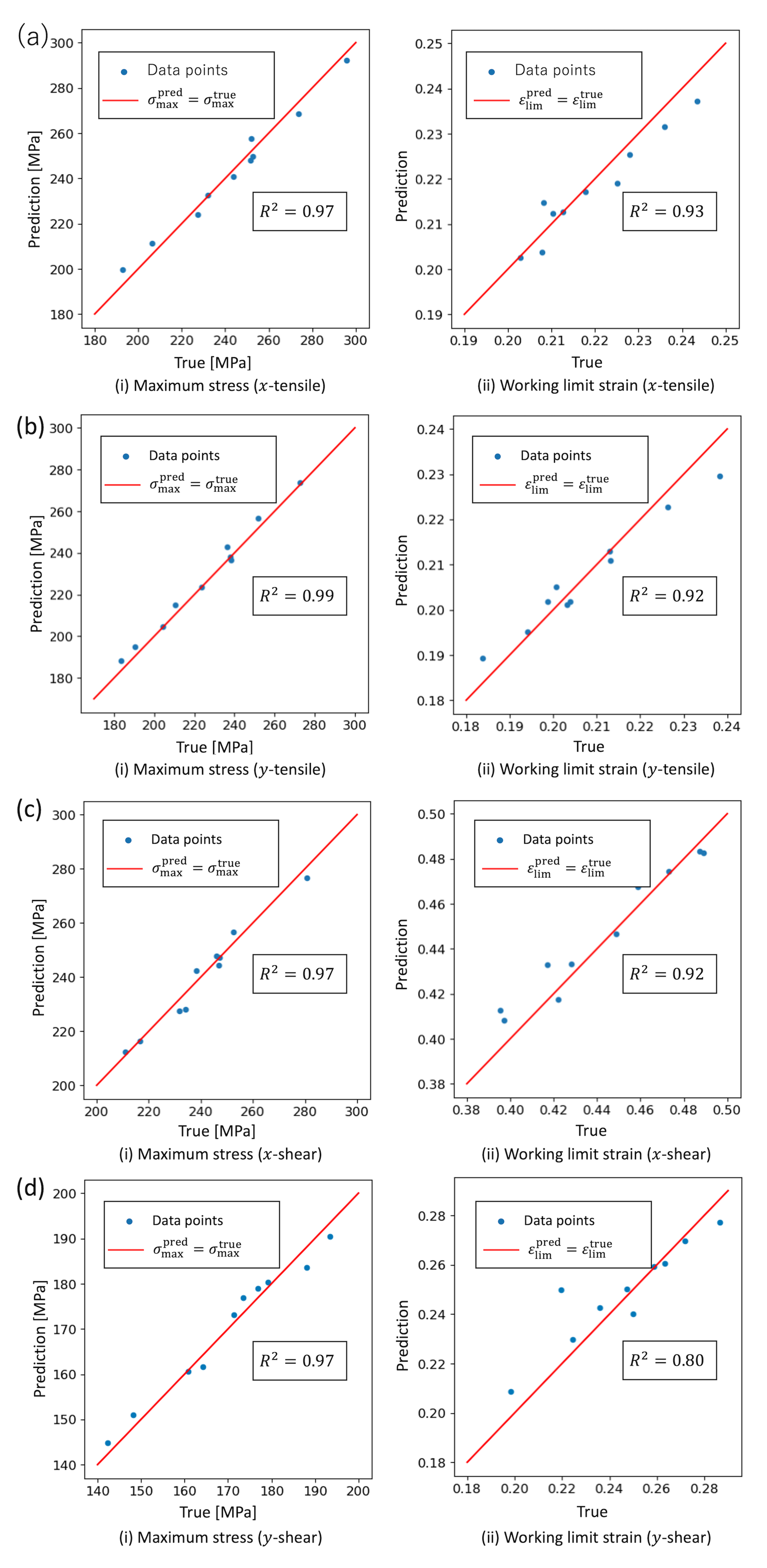}
\caption{Results of CNN. (a) Tensile toward $x$ direction. (b) Tensile toward $y$ direction. (c) Shear toward $x$ direction. (d) Shear toward $y$ direction. For each of (a) to (d), graph (i) shows results for maximum stress and graph (ii) shows results for working limit strain, respectively. Here, $R^2$ is the coefficient of determination.}
\label{Fig17}
\end{figure}

To confirm the generalization performance of CNN, the mechanical properties are predicted using the images of the DP steel microstructures not used for training.
The prediction results of CNN trained under the condition of tensile toward the $x$ direction for $10$ test data are shown in Fig. \ref{Fig17}a.
Here, the coefficient of determination for the prediction of $\sigma_{\mathrm{max}}$ is $R^2 = 0.97$ and that for $\varepsilon_{\mathrm{lim}}$ is $R^2 = 0.93$, indicating that the prediction of mechanical properties using CNN is highly accurate.
Similarly, the prediction results of CNN trained under the condition of tensile toward the $y$ direction, shear toward the $x$ direction, and shear toward the $y$ direction are shown in Figs. \ref{Fig17}b, c, and d respectively.
The coefficients of determination in predicting $\sigma_{\mathrm{max}}$ and $\varepsilon_{\mathrm{lim}}$ for tensile toward the $y$ direction are $R^2 = 0.99$ and $0.92$, those for shear toward the $x$ direction are $R^2 = 0.97$ and $0.92$, and those for shear toward the $y$ direction are $R^2 = 0.97$ and $0.80$, respectively, indicating that the prediction with high accuracy can be achieved by the trained CNN.

The coefficient of determination for the working limit strain $\varepsilon_{\mathrm{lim}}$ for shear toward the $y$ direction is smaller than those for the other deformation modes.
This can be explained by the features of the training data.
Figs. \ref{Fig13}, \ref{Fig14}, \ref{Fig15}, and \ref{Fig16} show the distribution of training data for each deformation mode.
For all deformation modes, the larger the fraction of martensite, the more scattered the distribution, and this tendency is more apparent in the case of shear toward the $y$ direction.
Comparisons of Fig. \ref{Fig13} with Fig. \ref{Fig15} and Fig. \ref{Fig14} with Fig. \ref{Fig16} show that shear deformation has more variability than tensile deformation.
Comparisons of Fig. \ref{Fig13} with Fig. \ref{Fig14} and Fig. \ref{Fig15} with Fig. \ref{Fig16} show that there is more variability in the deformation toward the $y$ direction than toward the $x$ direction.
This is because the initial state of phase-field analysis has an initial grain boundary parallel to the $x$ direction, as shown in Fig. \ref{Fig2}a.
Furthermore, Fig. \ref{Fig16} shows that for similar fractions of martensite, the maximum stress $\sigma_{\mathrm{max}}$ does not change much, but the working limit strain $\varepsilon_{\mathrm{lim}}$ does.
These are the reasons why the coefficient of determination for shear toward the $y$ direction is smaller than those for the other deformation modes.
Therefore, it can be said that the performance of CNN for shear toward the $y$ direction can be improved by adding training data.
In particular, it is necessary to add training data with large fractions of martensite.
\clearpage
\subsection{Investigation of DP steel microstructure with high strength and ductility}\label{subsec7}
An inverse analysis framework is constructed using the trained GAN and CNN.
Inverse analysis is performed to investigate DP steel microstructures that exhibit high strength and ductility.

\begin{figure}[t]%
\centering
\includegraphics[width=0.8\textwidth]{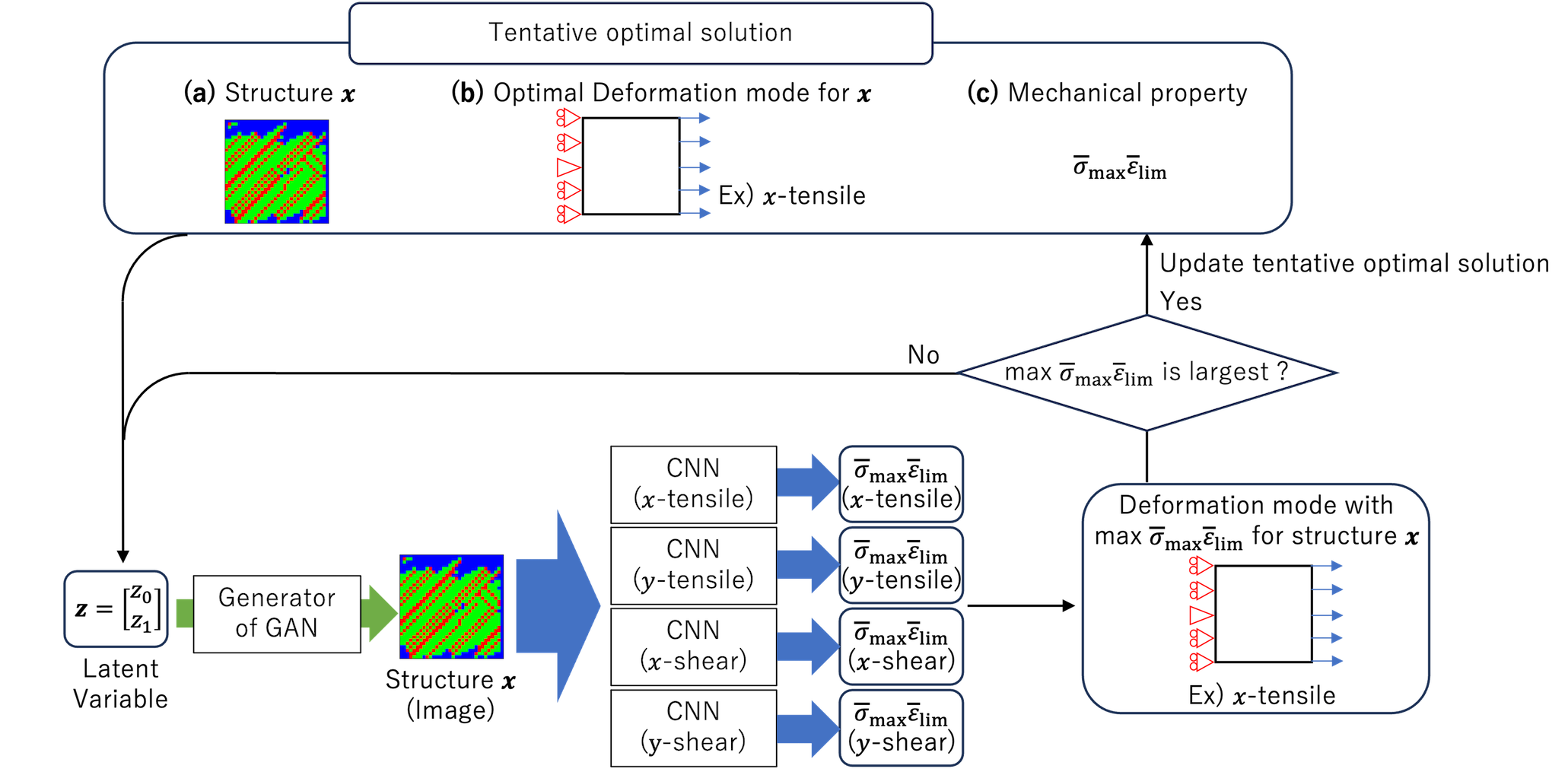}
\caption{Detailed process of the random search.}
\label{Fig18}
\end{figure}

The detailed process of the random search is shown in Fig. \ref{Fig18}.
One loop shown in Fig. \ref{Fig18} corresponds to one iteration.
At the beginning of the iteration, a two-dimensional vector $\boldsymbol{z}=[z_0,z_1]$ is randomly selected.
This $\boldsymbol{z}$ is the only input to GAN, and it is not necessary to provide the desired $\sigma_{\mathrm{max}}$ or $\varepsilon_{\mathrm{lim}}$ as other inputs.
Thus, one image of the DP steel microstructure is generated per an iteration.

CNN predicts $\bar{\sigma}_{\mathrm{max}}\bar{\varepsilon}_{\mathrm{lim}}$ for each of the four deformation modes.
That is, for one microstructure, four $\bar{\sigma}_{\mathrm{max}}\bar{\varepsilon}_{\mathrm{lim}}$ values are predicted as shown in Fig. \ref{Fig18}.
The deformation mode at the largest of the four $\bar{\sigma}_{\mathrm{max}}\bar{\varepsilon}_{\mathrm{lim}}$ values is the optimal deformation mode for the microstructure generated.
Now, when in the first iteration, there are the following three tentative optimal solutions: (a) the first is the microstructure, (b) the second is the optimal deformation mode for the microstructure (a), and (c) the third is $\bar{\sigma}_{\mathrm{max}}\bar{\varepsilon}_{\mathrm{lim}}$ under the optimal deformation mode (b) for the microstructure (a).
At the second and subsequent iterations, we update the tentative optimal solutions (a), (b), and (c) when the obtained (c) is the best among the iterations so far.

The iteration number of random searches is set as $5000$.
The random search stops when the specified number of iterations is reached.
Therefore, the number of iterations should be set sufficiently large.
Since mechanical properties generally do not show a one-to-one correspondence to structures, multiple solutions are possible in inverse analysis.
With the current method of using a single objective function such as the product of strength and ductility, there is a problem that only one solution is obtained when multiple solutions should be possible.
This problem can be avoided if the optimization is performed with the strength and ductility as the separate objective functions.
Among the Pareto optimal solutions obtained by multi-objective optimization, it is more practical to select the solution with higher strength when strength is important and the solution with higher ductility when ductility is important.
Depending on the desired balance between strength and ductility, different structures can practically be selected from the Pareto optimal solutions.

In the case of complex deformations, the speed and memory requirements of the proposed method are superior to the ones of the existing methods.
In this study, four deformation modes are considered simultaneously in one iteration, whereas only one deformation mode is considered in one iteration in the existing method \cite{hiraide2021application}.
If the four deformation modes are considered exhaustively, the existing method requires four times as many iterations as the proposed method.
The calculation speed is mostly determined by the iteration number.
Nevertheless, the memory requirement remains the same, the proposed method can be said to be four times faster than the existing method.
\begin{table}[t]
\caption{Comparison of mechanical properties of the proposed microstructure under shear toward $x$ direction}\label{Table6}
\centering
\begin{tabular}{l|l l l}
\hline
 & $\sigma_{\mathrm{max}}$  & $\varepsilon_{\mathrm{lim}}$ &  $\bar{\sigma}_{\mathrm{max}}\bar {\varepsilon}_{\mathrm{lim}}$\\ \hline
Results predicted by CNN & 261 MPa & 0.430 & 0.517\\ 
Results obtained from FEM & 260 MPa & 0.427 & 0.508\\ 
Relative error & 0.76\% & 0.25\% &1.69\% \\ 
\hline
\end{tabular}
\end{table}

\begin{figure}[t]%
\centering
\includegraphics[width=0.45\textwidth]{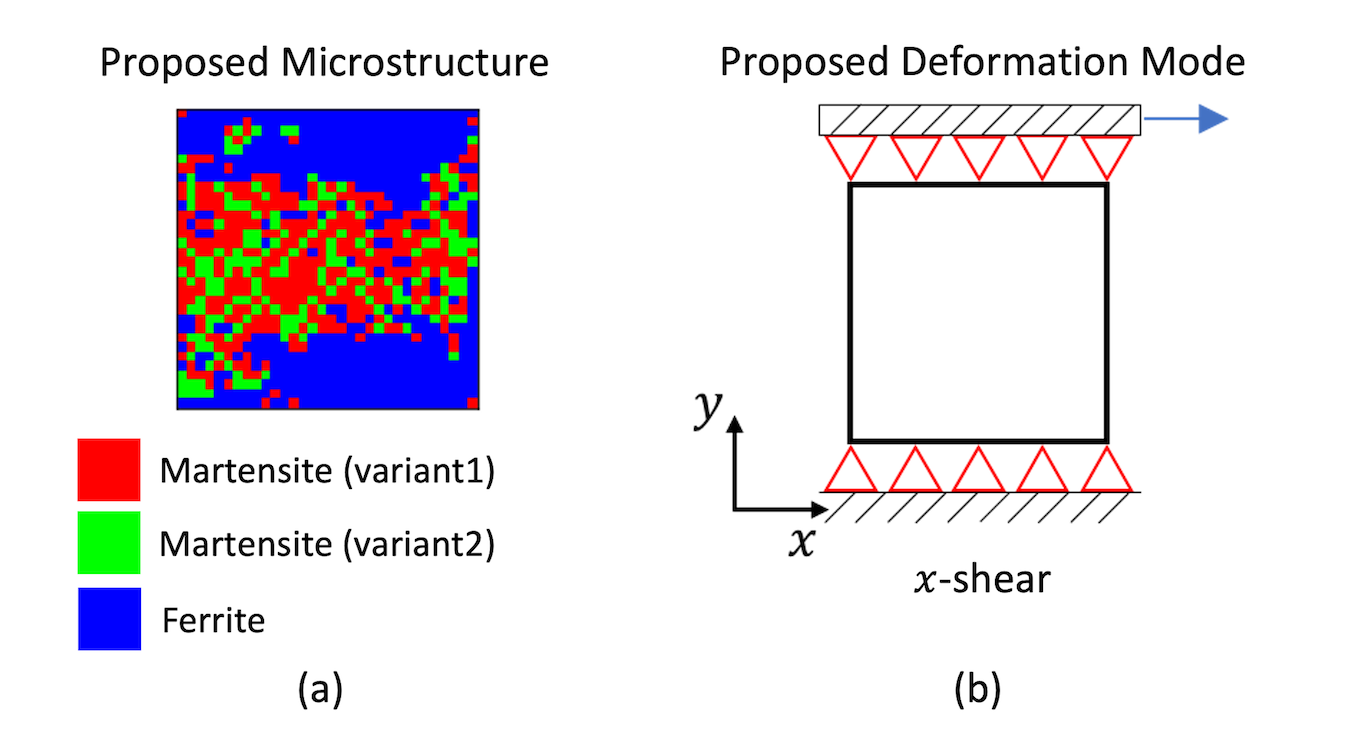}
\caption{Proposed microstructure and deformation mode. (a) Proposed DP steel microstructure. (b) Proposed deformation mode.}
\label{Fig19}
\end{figure}

The proposed microstructure and deformation mode are shown in Fig. \ref{Fig19}a and b, respectively. 
In other words, when the microstructure shown in Fig. \ref{Fig19}a is deformed by shear toward the $x$ direction, $\bar{\sigma}_{\mathrm{max}}\bar{\varepsilon}_{\mathrm{lim}}$ is the highest, indicating high strength and ductility.
The maximum stress $\sigma_{\mathrm{max}}$ of the proposed microstructure is $\SI{261}{\ \MPa}$, the working limit strain $\varepsilon_{\mathrm{lim}}$ is $0.430$, and $\bar{\sigma}_{\mathrm{max}}\bar{\varepsilon}_{\mathrm{lim}}$ is $0.517$.
These values are predicted by CNN during inverse analysis.
Here, FEM is performed for the proposed microstructure shown in Fig. \ref{Fig19}a.
As a result, the maximum stress $\sigma_{\mathrm{max}}$ of the proposed DP steel microstructure is determined to be $\SI{260}{\ \MPa}$, the working limit strain $\varepsilon_{\mathrm{lim}}$ is $0.427$, and $\bar{\sigma}_{\mathrm {max}}\bar{\varepsilon}_{\mathrm{lim}}$ is $0.508$.

Table \ref{Table6} shows a summary of the mechanical properties predicted by CNN and obtained by the FEM analysis when the proposed microstructure is deformed by the proposed deformation mode (shear toward $x$ direction).
The relative errors of $\sigma_{\mathrm{max}}$, $\varepsilon_{\mathrm{lim}}$, and $\bar{\sigma}_{\mathrm{max}}\bar {\varepsilon}_{\mathrm{lim}}$ are $0.76\%$, $0.25\%$, and $1.69\%$, respectively, indicating that the inverse analysis is performed with high accuracy.

\begin{figure}[t]%
\centering
\includegraphics[width=0.45\textwidth]{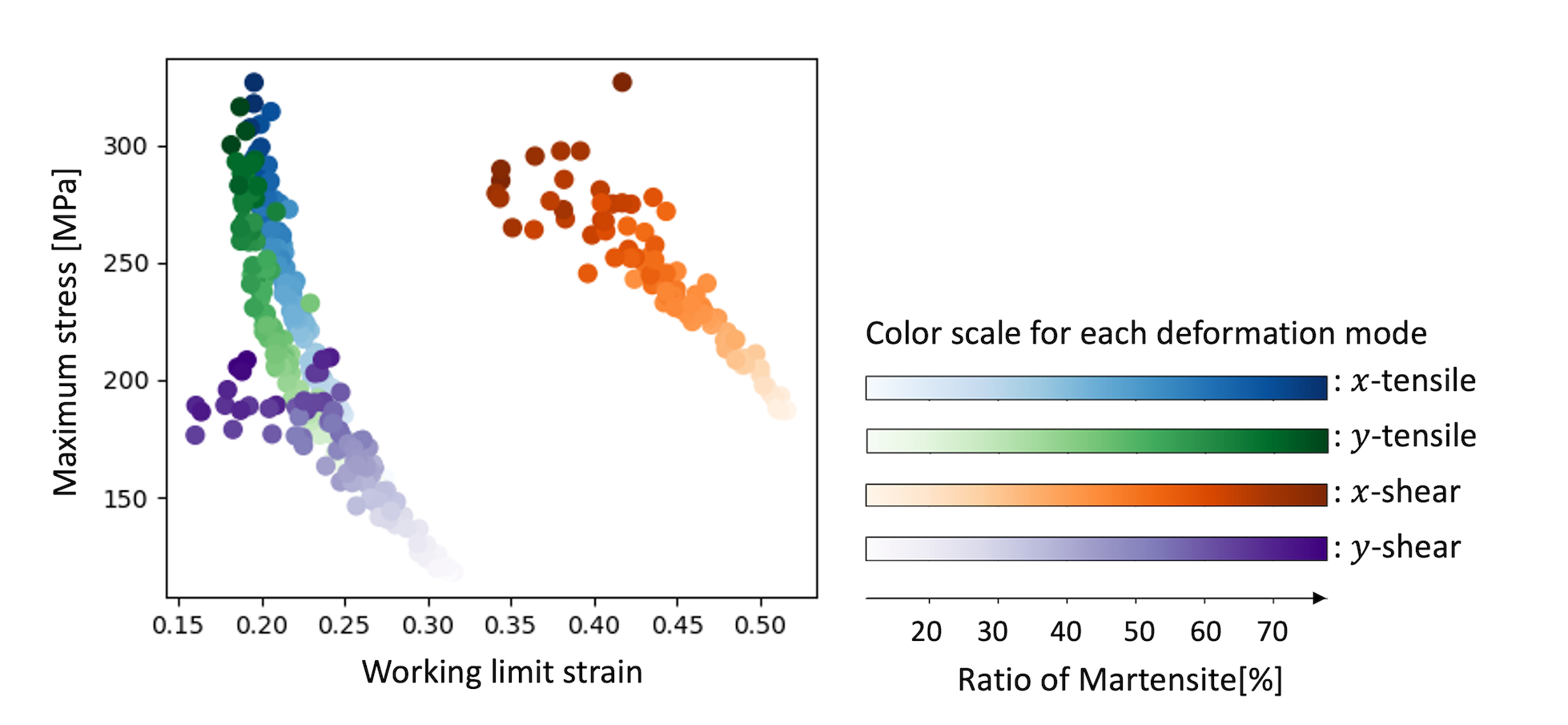}
\caption{Relationship between maximum stress and working limit strain.}
\label{Fig20}
\end{figure}

Fig.  \ref{Fig20} shows the relationship between the maximum stress $\sigma_{\mathrm{max}}$ and the working limit strain $\varepsilon_{\mathrm{lim}}$.
The values of $\bar{\sigma}_{\mathrm{max}}\bar {\varepsilon}_{\mathrm{lim}}$ for shear toward the $x$ direction are located in the upper right region of the figure, which means high strength and ductility. 
This indicates that the proposed deformation mode is reasonable.

\begin{figure}[t]%
\centering
\includegraphics[width=0.45\textwidth]{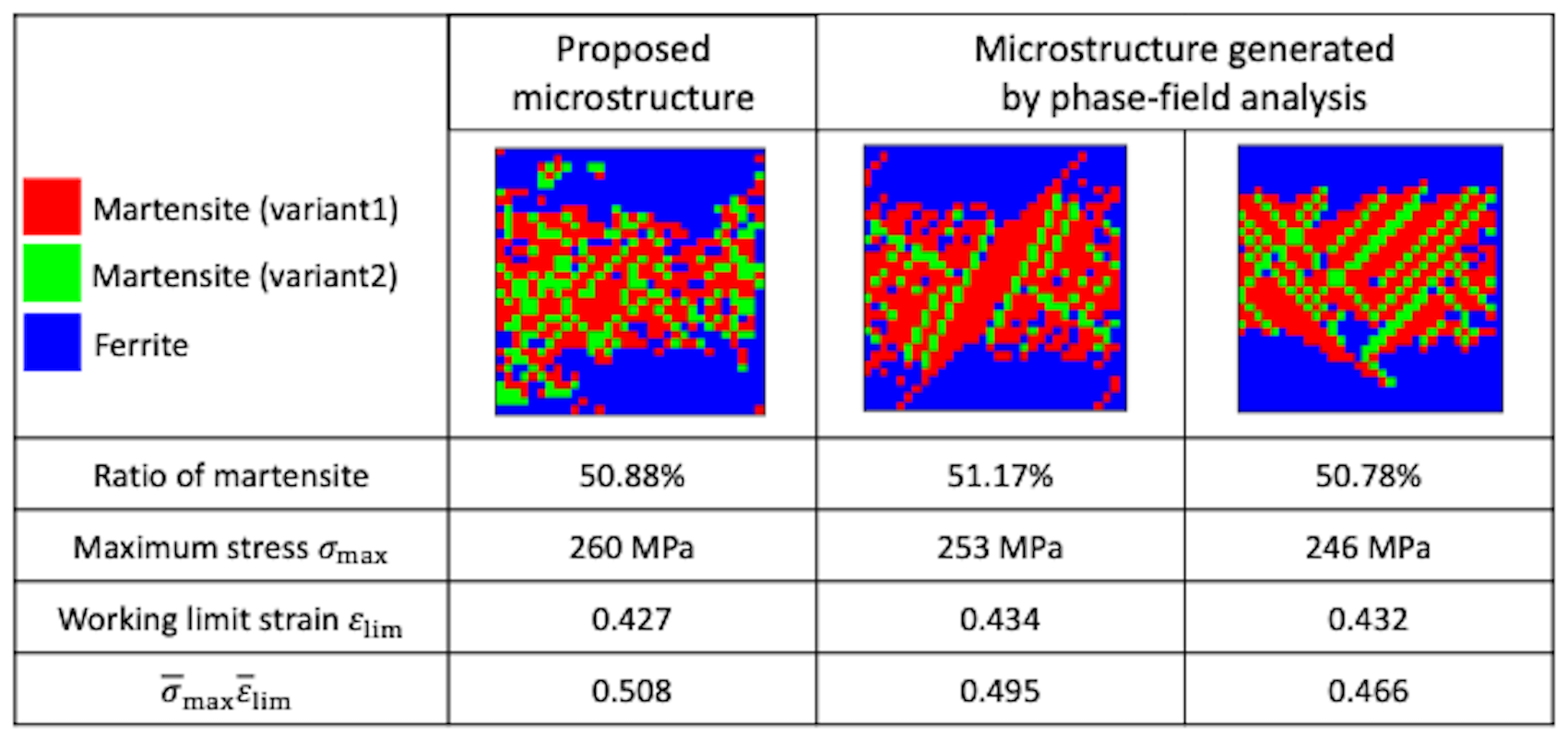}
\caption{Comparison of the microstructure proposed by the present framework and the microstructures obtained by phase-field analysis.}
\label{Fig21}
\end{figure}

Fig. \ref{Fig21} shows a comparison of the proposed microstructure and the microstructures obtained by phase-field analysis.
The microstructures compared in Fig. \ref{Fig21} have similar fractions to martensite.
The proposed microstructure has a finer distribution of martensite variant$1$ and variant$2$ than the microstructure obtained by phase-field analysis.
In other words, the grain size of martensite in the proposed microstructure is smaller than that obtained by phase-field analysis.
The mechanical properties are compared among three DP steel microstructures in Fig. \ref{Fig21} when they are deformed by shear toward the $x$ direction.
The results show that the $\bar{\sigma}_{\mathrm{max}}\bar{\varepsilon}_{\mathrm{lim}}$ of the proposed DP steel structure is the highest.
This is because $\sigma_{\mathrm{max}}$ is higher in the proposed microstructure, whereas $\varepsilon_{\mathrm{lim}}$ is lower, but $\sigma_{\mathrm{max}}$ has a greater effect.
This characteristics is shown in polycrystalline materials with small grain size \cite{tsuji2008managing}, and it is reasonable that such a DP steel microstructure is proposed.

Although in this study we focus on microstructures near grain boundaries, the scale of the proposed structures would be sufficient for practical use when larger analysis domains are used for this framework.
For further practicality, not only structures but also process parameters that can be controlled when the structures are created experimentally should be proposed.
Experimental data are required to propose process parameters, but it is difficult to obtain sufficient data sets.
To solve this problem, a technique that combines experimental and simulation data is used.
The framework in this study is suitable for this technique because it is based on the phase-field method and dislocation-crystal plasticity FEM, which are physical simulation methods that reproduce experiments.

\subsection{Discussion of employing random search for inverse analysis}\label{subsec8}
\begin{figure}[t]%
\centering
\includegraphics[width=0.6\textwidth]{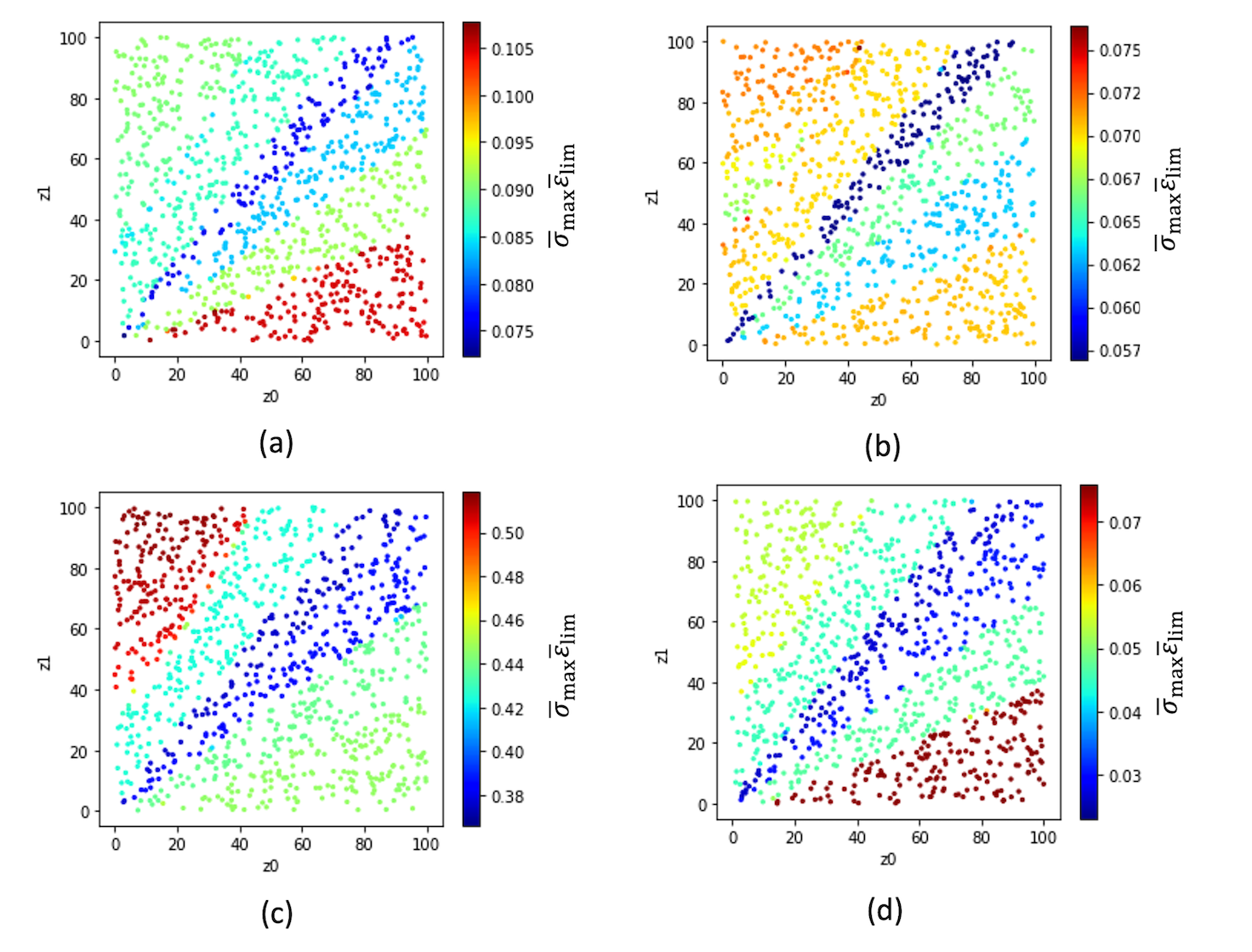}
\caption{Distribution of the product of the normalized maximum stress and the working limit strain in the latent variable space. (a) Tensile toward $x$ direction. (b) Tensile toward $y$ direction. (c) Shear toward $x$ direction. (d) Shear toward $y$ direction.}
\label{Fig22}
\end{figure}

Here, the appropriateness of employing the random search for the inverse analysis is considered by comparing it with the gradient descent method.
For the purposes of this examination, the relationship between the latent variables and the product of the normalized maximum stress and the working limit strain, $\bar{\sigma}_{\mathrm{max}}\bar{\varepsilon}_{\mathrm{lim}}$, is shown in Fig. \ref{Fig22}.
Fig. \ref{Fig22}a shows the case of tensile toward the $x$ direction.
In this figure, the region where  $\bar{\sigma}_{\mathrm{max}}\bar{\varepsilon}_{\mathrm{lim}}$ becomes high is divided into two parts by the blue plots at around $z_0=z_1$.
The global optimal solution should be taken in the lower right region where $z_0$ is high and $z_1$ is low.
However, if the initial value is taken in the upper left region where $z_0$ is low and $z_1$ is high, the optimal solution will fall into the local solution near the initial value.

Similarly, for the tensile toward the $y$ direction shown in Fig. \ref{Fig22}b, the shear toward the $x$ direction shown in Fig. \ref{Fig22}c, and the shear toward the $y$ direction shown in Fig. \ref{Fig22}d, the optimal solutions fall into local ones, depending on their initial values.
For this reason, in this study, we employ random search, which enables an exhaustive search of the latent variable space.
The computational cost, which is a concern in the use of random search, is not considered to be a problem.
This is because the high-dimensional images can be dropped into a low-dimensional latent variable space.

\begin{figure}[t]%
\centering
\includegraphics[width=0.3\textwidth]{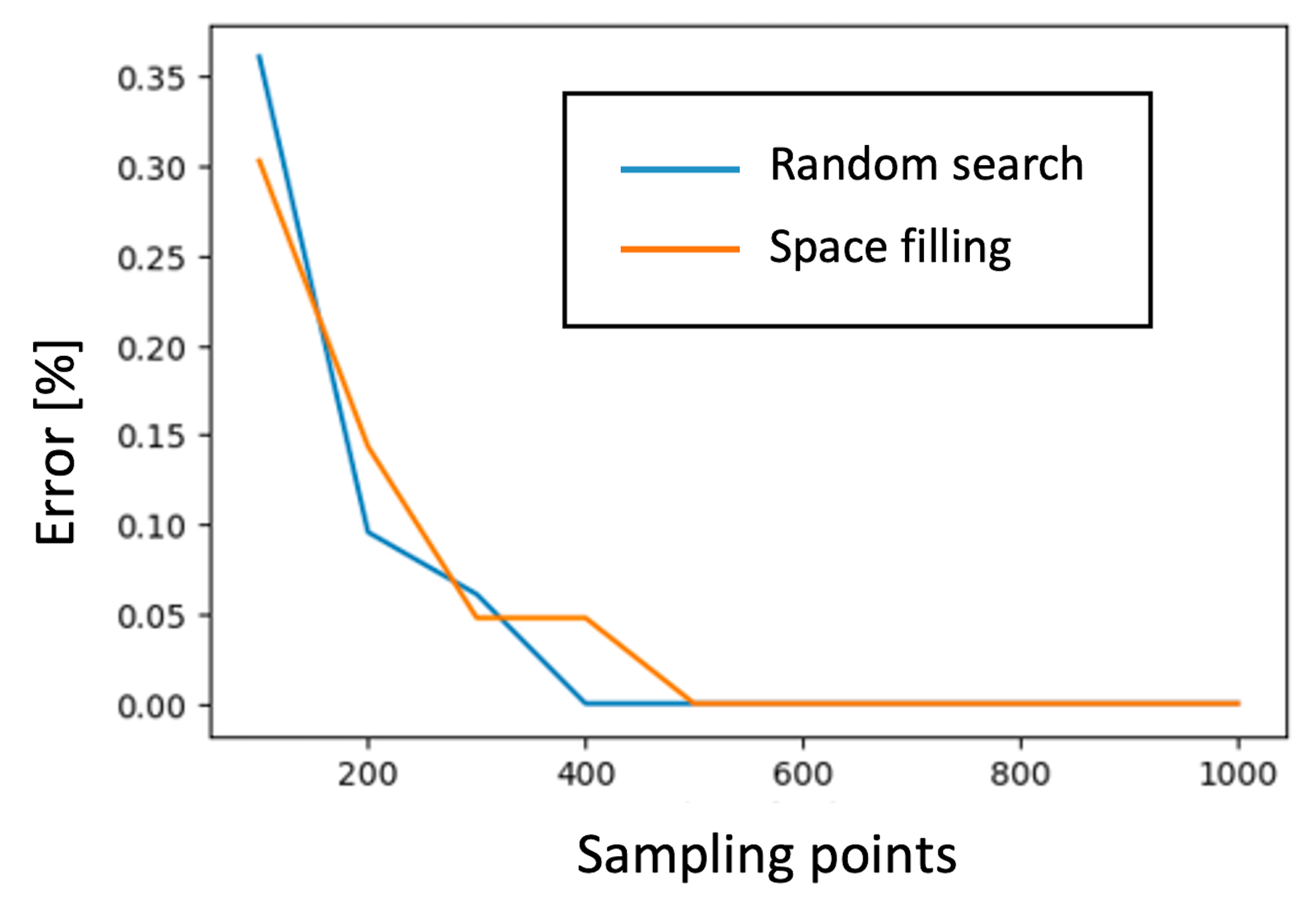}
\caption{Comparison of errors between random search and space-filling design.}
\label{Fig23}
\end{figure}

We consider methods of sampling the space that are not based on gradients.
We focus here on the space-filling design, which is commonly used to select design variables for experiments.
The space-filling design is a kind of design of experiments that involves uniform sampling within the design space.
Recently, it has been used to efficiently obtain large numbers of training data for machine learning \cite{bessa2017framework}.
In this study, the latent variable space is sampled on the basis of the space-filling design and compared with that obtained by a random search.

Fig. \ref{Fig23} shows the relationship between the number of sampling points, which is called the number of iterations in the random search, and the error of $\bar{\sigma}_{\mathrm{max}}\bar{\varepsilon}_{\mathrm{lim}}$.
The error is the difference from $\bar{\sigma}_{\mathrm{max}}\bar{\varepsilon}_{\mathrm{lim}}$ obtained by the random search for a sufficiently large number of sampling points, $5000$.
The number of sampling points is set from $100$ to $1000$ in increments of $100$. 
At each number of sampling points, ten searches are performed, and the average of the errors is plotted.
Fig. \ref{Fig23} shows that the space-filling design is not markedly superior to the random search.
This is because in the a low-dimensional space, the solution can be reached with a realistically feasible amount of computation even if the random search is used \cite{hiraide2023inverse}.
Since the space-filling design is naturally more reasonable than the random search, in which sampling points are very close to each other, it would be more effective when the search space is a higher-dimensional one.

\section{Conclusion}\label{sec6}
In this study, we developed an inverse analysis framework that can propose an optimal microstructure of DP steel using machine learning. 
The developed inverse analysis framework used the combination of GAN, which generates microstructures, and CNN, which predicts the maximum stress and working limit strain from DP steel microstructures.

The novelty of this study as a framework is twofold.
The first is that microstructures can be proposed by considering the trade-off of DP steels without specifying the desired mechanical properties in advance.
The second is that microstructures are proposed by considering four deformation modes simultaneously.
Compared with the case that four deformation modes, the proposed method requires approximately one-fourth of iteration number in existing methods.
Due to the reduction of the iteration number, the speed is significantly improved from the existing method keeping the same memory requirement.
In addition, the novelty as a method of material exploration of DP steels is that the proposed microstructures are based on the phase-field method, and therefore, the results are consistent with the trends observed in experiments.

As a result of implementing the developed inverse analysis framework, the following were confirmed.
\begin{enumerate}[1.]
\item GAN could generate images with features of DP steel microstructures, which were generated by the phase-field analysis.
\item CNN could predict maximum stress and working limit strain, which were obtained by dislocation-crystal plasticity FEM.
\item The product of the normalized maximum stress and the working limit strain was used as the criterion for high strength and ductility, and by performing random search, the optimal DP steel microstructures and deformation mode were proposed.
\item The martensitic phase of the proposed microstructure showed fine grains, which was consistent with the trend observed in the experiments.
\end{enumerate}

The inverse analysis framework developed in this study has the following advantages:
The first is that GAN generates DP steel microstructures from the results of phase-field analysis; thus, the images generated by GAN have the features of actual DP steel.
The second is that it is easy to adapt the inverse analysis framework to other mechanical properties because when it is necessary to change the target mechanical properties, for example, changing the maximum stress to yield stress, CNN only needs to be trained again.

\section*{Acknowledgements}
The authors greatly appreciate financial support by the JST CREST Nanomechanics (JPMJCR1994).

\section*{Data Availability}
Data will be made available on request.



\bibliographystyle{elsarticle-num} 
\bibliography{reference.bib}

\begin{thebibliography}{10}
\expandafter\ifx\csname url\endcsname\relax
  \def\url#1{\texttt{#1}}\fi
\expandafter\ifx\csname urlprefix\endcsname\relax\def\urlprefix{URL }\fi
\expandafter\ifx\csname href\endcsname\relax
  \def\href#1#2{#2} \def\path#1{#1}\fi

\bibitem{ritchie2011conflicts}
R.~O. Ritchie, The conflicts between strength and toughness, Nat Mater 10~(11) (2011) 817--822.
\newblock \href {https://doi.org/10.1038/nmat3115} {\path{doi:10.1038/nmat3115}}.

\bibitem{tasan2015overview}
C.~C. Tasan, M.~Diehl, D.~Yan, M.~Bechtold, F.~Roters, L.~Schemmann, C.~Zheng, N.~Peranio, D.~Ponge, M.~Koyama, K.~Tsuzaki, D.~Raabe, An overview of dual-phase steels: advances in microstructure-oriented processing and micromechanically guided design, Annu Rev Mater Res. 45 (2015) 391--431.
\newblock \href {https://doi.org/10.1146/annurev-matsci-070214-021103} {\path{doi:10.1146/annurev-matsci-070214-021103}}.

\bibitem{park2014effect}
K.~Park, M.~Nishiyama, N.~Nakada, T.~Tsuchiyama, S.~Takaki, Effect of the martensite distribution on the strain hardening and ductile fracture behaviors in dual-phase steel, Mater Sci Eng A 604 (2014) 135--141.
\newblock \href {https://doi.org/10.1016/j.msea.2014.02.058} {\path{doi:10.1016/j.msea.2014.02.058}}.

\bibitem{calcagnotto2011deformation}
M.~Calcagnotto, Y.~Adachi, D.~Ponge, D.~Raabe, Deformation and fracture mechanisms in fine-and ultrafine-grained ferrite/martensite dual-phase steels and the effect of aging, Acta Mater 59~(2) (2011) 658--670.
\newblock \href {https://doi.org/10.1016/j.actamat.2010.10.002} {\path{doi:10.1016/j.actamat.2010.10.002}}.

\bibitem{son2005ultrafine}
Y.~I. Son, Y.~K. Lee, K.-T. Park, C.~S. Lee, D.~H. Shin, Ultrafine grained ferrite--martensite dual phase steels fabricated via equal channel angular pressing: microstructure and tensile properties, Acta Mater 53~(11) (2005) 3125--3134.
\newblock \href {https://doi.org/10.1016/j.actamat.2005.02.015} {\path{doi:10.1016/j.actamat.2005.02.015}}.

\bibitem{TERADA20002285}
K.~Terada, M.~Hori, T.~Kyoya, N.~Kikuchi, Simulation of the multi-scale convergence in computational homogenization approaches, Int J Solids Struct 37~(16) (2000) 2285--2311.
\newblock \href {https://doi.org/10.1016/S0020-7683(98)00341-2} {\path{doi:10.1016/S0020-7683(98)00341-2}}.

\bibitem{BAI2020105735}
L.~Bai, C.~Gong, X.~Chen, Y.~Sun, L.~Xin, H.~Pu, Y.~Peng, J.~Luo, Mechanical properties and energy absorption capabilities of functionally graded lattice structures: Experiments and simulations, Int J Mech Sci 182 (2020) 105735.
\newblock \href {https://doi.org/10.1016/j.ijmecsci.2020.105735} {\path{doi:10.1016/j.ijmecsci.2020.105735}}.

\bibitem{WANG2023103948}
Z.~Wang, W.~Chen, H.~Hao, Y.~Dong, Z.~Huang, Numerical prediction of blast fragmentation of reinforced concrete slab using ale-fem-sph coupling method, Finite Elem Anal Des 220 (2023) 103948.
\newblock \href {https://doi.org/10.1016/j.finel.2023.103948} {\path{doi:10.1016/j.finel.2023.103948}}.

\bibitem{YAP2023107771}
Y.~L. Yap, W.~Toh, A.~Giam, F.~R. Yong, K.~I. Chan, J.~W.~S. Tay, S.~S. Teong, R.~Lin, T.~Y. Ng, Topology optimization and 3d printing of micro-drone: Numerical design with experimental testing, Int J Mech Sci 237 (2023) 107771.
\newblock \href {https://doi.org/10.1016/j.ijmecsci.2022.107771} {\path{doi:10.1016/j.ijmecsci.2022.107771}}.

\bibitem{yeddu2012three}
H.~K. Yeddu, A.~Malik, J.~{\AA}gren, G.~Amberg, A.~Borgenstam, Three-dimensional phase-field modeling of martensitic microstructure evolution in steels, Acta Mater 60~(4) (2012) 1538--1547.
\newblock \href {https://doi.org/10.1016/j.actamat.2011.11.039} {\path{doi:10.1016/j.actamat.2011.11.039}}.

\bibitem{militzer2006three}
M.~Militzer, M.~Mecozzi, J.~Sietsma, S.~Van~der Zwaag, Three-dimensional phase field modelling of the austenite-to-ferrite transformation, Acta Mater 54~(15) (2006) 3961--3972.
\newblock \href {https://doi.org/10.1016/j.actamat.2006.04.029} {\path{doi:10.1016/j.actamat.2006.04.029}}.

\bibitem{TAKAHAMA20122916}
Y.~Takahama, M.~Santofimia, M.~Mecozzi, L.~Zhao, J.~Sietsma, Phase field simulation of the carbon redistribution during the quenching and partitioning process in a low-carbon steel, Acta Mater 60~(6) (2012) 2916--2926.
\newblock \href {https://doi.org/10.1016/j.actamat.2012.01.055} {\path{doi:10.1016/j.actamat.2012.01.055}}.

\bibitem{MECOZZI2016245}
M.~Mecozzi, J.~Eiken, M.~Santofimia, J.~Sietsma, Phase field modelling of microstructural evolution during the quenching and partitioning treatment in low-alloy steels, Comput Mater Sci 112 (2016) 245--256.
\newblock \href {https://doi.org/10.1016/j.commatsci.2015.10.048} {\path{doi:10.1016/j.commatsci.2015.10.048}}.

\bibitem{woo2012stress}
W.~Woo, V.~Em, E.-Y. Kim, S.~Han, Y.~Han, S.-H. Choi, Stress--strain relationship between ferrite and martensite in a dual-phase steel studied by in situ neutron diffraction and crystal plasticity theories, Acta Mater 60~(20) (2012) 6972--6981.
\newblock \href {https://doi.org/10.1016/j.actamat.2012.08.054} {\path{doi:10.1016/j.actamat.2012.08.054}}.

\bibitem{kadkhodapour2011micro}
J.~Kadkhodapour, A.~Butz, S.~Ziaei-Rad, S.~Schmauder, A micro mechanical study on failure initiation of dual phase steels under tension using single crystal plasticity model, Int J Plast 27~(7) (2011) 1103--1125.
\newblock \href {https://doi.org/10.1016/j.ijplas.2010.12.001} {\path{doi:10.1016/j.ijplas.2010.12.001}}.

\bibitem{ZAAFARANI20061863}
N.~Zaafarani, D.~Raabe, R.~Singh, F.~Roters, S.~Zaefferer, Three-dimensional investigation of the texture and microstructure below a nanoindent in a cu single crystal using 3d ebsd and crystal plasticity finite element simulations, Acta Mater 54~(7) (2006) 1863--1876.
\newblock \href {https://doi.org/10.1016/j.actamat.2005.12.014} {\path{doi:10.1016/j.actamat.2005.12.014}}.

\bibitem{LU2020102703}
X.~Lu, J.~Zhao, Z.~Wang, B.~Gan, J.~Zhao, G.~Kang, X.~Zhang, Crystal plasticity finite element analysis of gradient nanostructured twip steel, Int J Plast 130 (2020) 102703.
\newblock \href {https://doi.org/10.1016/j.ijplas.2020.102703} {\path{doi:10.1016/j.ijplas.2020.102703}}.

\bibitem{LIAO2020105685}
D.~Liao, S.-P. Zhu, B.~Keshtegar, G.~Qian, Q.~Wang, Probabilistic framework for fatigue life assessment of notched components under size effects, Int J Mech Sci 181 (2020) 105685.
\newblock \href {https://doi.org/10.1016/j.ijmecsci.2020.105685} {\path{doi:10.1016/j.ijmecsci.2020.105685}}.

\bibitem{shinoda2004rapid}
W.~Shinoda, M.~Shiga, M.~Mikami, Rapid estimation of elastic constants by molecular dynamics simulation under constant stress, Phys Rev B 69~(13) (2004) 134103.
\newblock \href {https://doi.org/10.1103/PhysRevB.69.134103} {\path{doi:10.1103/PhysRevB.69.134103}}.

\bibitem{LOPES2021106650}
D.~Lopes, R.~Agujetas, H.~Puga, J.~Teixeira, R.~Lima, J.~Alejo, C.~Ferrera, Analysis of finite element and finite volume methods for fluid-structure interaction simulation of blood flow in a real stenosed artery, Int J Mech Sci 207 (2021) 106650.
\newblock \href {https://doi.org/10.1016/j.ijmecsci.2021.106650} {\path{doi:10.1016/j.ijmecsci.2021.106650}}.

\bibitem{NUTARO2023111990}
J.~Nutaro, B.~Stump, P.~Shukla, Discrete event cellular automata: A new approach to cellular automata for computational material science, Comput Mater Sci 219 (2023) 111990.
\newblock \href {https://doi.org/10.1016/j.commatsci.2022.111990} {\path{doi:10.1016/j.commatsci.2022.111990}}.

\bibitem{bock2019review}
F.~E. Bock, R.~C. Aydin, C.~J. Cyron, N.~Huber, S.~R. Kalidindi, B.~Klusemann, A review of the application of machine learning and data mining approaches in continuum materials mechanics, Front Mater Sci 6 (2019) 110.
\newblock \href {https://doi.org/10.3389/fmats.2019.00110} {\path{doi:10.3389/fmats.2019.00110}}.

\bibitem{ramprasad2017machine}
R.~Ramprasad, R.~Batra, G.~Pilania, A.~Mannodi-Kanakkithodi, C.~Kim, Machine learning in materials informatics: recent applications and prospects, Npj Comput Mater 3~(1) (2017) 1--13.
\newblock \href {https://doi.org/10.1038/s41524-017-0056-5} {\path{doi:10.1038/s41524-017-0056-5}}.

\bibitem{gubernatis2018machine}
J.~Gubernatis, T.~Lookman, Machine learning in materials design and discovery: Examples from the present and suggestions for the future, Phys Rev Mater 2~(12) (2018) 120301.
\newblock \href {https://doi.org/10.1103/PhysRevMaterials.2.120301} {\path{doi:10.1103/PhysRevMaterials.2.120301}}.

\bibitem{butler2018machine}
K.~T. Butler, D.~W. Davies, H.~Cartwright, O.~Isayev, A.~Walsh, Machine learning for molecular and materials science, Nature 559~(7715) (2018) 547--555.
\newblock \href {https://doi.org/10.1038/s41586-018-0337-2} {\path{doi:10.1038/s41586-018-0337-2}}.

\bibitem{liu2019predicting}
Y.~Liu, J.~Wu, G.~Yang, T.~Zhao, S.~Shi, Predicting the onset temperature (tg) of gexse1- x glass transition: a feature selection based two-stage support vector regression method, Sci Bull 64~(16) (2019) 1195--1203.
\newblock \href {https://doi.org/10.1016/j.scib.2019.06.026} {\path{doi:10.1016/j.scib.2019.06.026}}.

\bibitem{xie2018crystal}
T.~Xie, J.~C. Grossman, Crystal graph convolutional neural networks for an accurate and interpretable prediction of material properties, Phys Rev Lett 120~(14) (2018) 145301.
\newblock \href {https://doi.org/10.1103/PhysRevLett.120.145301} {\path{doi:10.1103/PhysRevLett.120.145301}}.

\bibitem{DUAN2013524}
Z.~Duan, S.~Kou, C.~Poon, Using artificial neural networks for predicting the elastic modulus of recycled aggregate concrete, Constr Build Mater 44 (2013) 524--532.
\newblock \href {https://doi.org/10.1016/j.conbuildmat.2013.02.064} {\path{doi:10.1016/j.conbuildmat.2013.02.064}}.

\bibitem{yamanaka2020deep}
A.~Yamanaka, R.~Kamijyo, K.~Koenuma, I.~Watanabe, T.~Kuwabara, Deep neural network approach to estimate biaxial stress-strain curves of sheet metals, Mater Des 559~(7715) (2018) 547--555.
\newblock \href {https://doi.org/10.1016/j.matdes.2020.108970} {\path{doi:10.1016/j.matdes.2020.108970}}.

\bibitem{kalina2022automated}
K.~A. Kalina, L.~Linden, J.~Brummund, P.~Metsch, M.~K{\"a}stner, Automated constitutive modeling of isotropic hyperelasticity based on artificial neural networks, Comput Mech 69 (2022) 1--20.
\newblock \href {https://doi.org/10.1007/s00466-021-02090-6} {\path{doi:10.1007/s00466-021-02090-6}}.

\bibitem{gong2022additive}
X.~Gong, D.~Zeng, W.~Groeneveld-Meijer, G.~Manogharan, Additive manufacturing: A machine learning model of process-structure-property linkages for machining behavior of ti-6al-4v, Mater Sci Add Manuf 1~(6) (2022) 1--16.
\newblock \href {https://doi.org/10.18063/msam.v1i1.6} {\path{doi:10.18063/msam.v1i1.6}}.

\bibitem{kouraytem2021modeling}
N.~Kouraytem, X.~Li, W.~Tan, B.~Kappes, A.~D. Spear, Modeling process--structure--property relationships in metal additive manufacturing: a review on physics-driven versus data-driven approaches, J Phys Mater 4~(3) (2021) 032002.
\newblock \href {https://doi.org/10.1088/2515-7639/abca7b} {\path{doi:10.1088/2515-7639/abca7b}}.

\bibitem{RUIZ2022106785}
E.~Ruiz, D.~Ferreño, M.~Cuartas, B.~Arroyo, I.~A. Carrascal, I.~Rivas, F.~Gutiérrez-Solana, Application of machine learning algorithms for the optimization of the fabrication process of steel springs to improve their fatigue performance, Int J Fatigue 159 (2022) 106785.
\newblock \href {https://doi.org/10.1103/PhysRevMaterials.3.095201} {\path{doi:10.1103/PhysRevMaterials.3.095201}}.

\bibitem{zhu2021machine}
Q.~Zhu, Z.~Liu, J.~Yan, Machine learning for metal additive manufacturing: predicting temperature and melt pool fluid dynamics using physics-informed neural networks, Comput Mech 67 (2021) 619--635.
\newblock \href {https://doi.org/10.1007/s00466-020-01952-9} {\path{doi:10.1007/s00466-020-01952-9}}.

\bibitem{YANG2018278}
Z.~Yang, Y.~C. Yabansu, R.~Al-Bahrani, W.~keng Liao, A.~N. Choudhary, S.~R. Kalidindi, A.~Agrawal, Deep learning approaches for mining structure-property linkages in high contrast composites from simulation datasets, Comput Mater Sci 151 (2018) 278--287.
\newblock \href {https://doi.org/10.1016/j.commatsci.2018.05.014} {\path{doi:10.1016/j.commatsci.2018.05.014}}.

\bibitem{lu2019data}
X.~Lu, D.~G. Giovanis, J.~Yvonnet, V.~Papadopoulos, F.~Detrez, J.~Bai, A data-driven computational homogenization method based on neural networks for the nonlinear anisotropic electrical response of graphene/polymer nanocomposites, Comput Mech 64 (2019) 307--321.
\newblock \href {https://doi.org/10.1007/s00466-018-1643-0} {\path{doi:10.1007/s00466-018-1643-0}}.

\bibitem{krokos2022bayesian}
V.~Krokos, V.~Bui~Xuan, S.~P. Bordas, P.~Young, P.~Kerfriden, A bayesian multiscale cnn framework to predict local stress fields in structures with microscale features, Comput Mech 69~(3) (2022) 733--766.
\newblock \href {https://doi.org/10.1007/s00466-021-02112-3} {\path{doi:10.1007/s00466-021-02112-3}}.

\bibitem{li2019clustering}
H.~Li, O.~L. Kafka, J.~Gao, C.~Yu, Y.~Nie, L.~Zhang, M.~Tajdari, S.~Tang, X.~Guo, G.~Li, S.~Tang, G.~Cheng, W.~K. Liu, Clustering discretization methods for generation of material performance databases in machine learning and design optimization, Comput Mech 64 (2019) 281--305.
\newblock \href {https://doi.org/10.1007/s00466-019-01716-0} {\path{doi:10.1007/s00466-019-01716-0}}.

\bibitem{eidel2023deep}
B.~Eidel, Deep cnns as universal predictors of elasticity tensors in homogenization, Comput Methods Appl Mech Eng 403 (2023) 115741.
\newblock \href {https://doi.org/10.1016/j.cma.2022.115741} {\path{doi:10.1016/j.cma.2022.115741}}.

\bibitem{marshall2021autonomous}
A.~Marshall, S.~R. Kalidindi, Autonomous development of a machine-learning model for the plastic response of two-phase composites from micromechanical finite element models, JOM 73~(7) (2021) 2085--2095.
\newblock \href {https://doi.org/10.1007/s11837-021-04696-w} {\path{doi:10.1007/s11837-021-04696-w}}.

\bibitem{li2019machine}
X.~Li, C.~C. Roth, D.~Mohr, Machine-learning based temperature-and rate-dependent plasticity model: Application to analysis of fracture experiments on dp steel, Int J Plast 118 (2019) 320--344.
\newblock \href {https://doi.org/10.1016/j.ijplas.2019.02.012} {\path{doi:10.1016/j.ijplas.2019.02.012}}.

\bibitem{khosravani2017development}
A.~Khosravani, A.~Cecen, S.~R. Kalidindi, Development of high throughput assays for establishing process-structure-property linkages in multiphase polycrystalline metals: Application to dual-phase steels, Acta Mater 123 (2017) 55--69.
\newblock \href {https://doi.org/10.1016/j.actamat.2016.10.033} {\path{doi:10.1016/j.actamat.2016.10.033}}.

\bibitem{MARTINEZOSTORMUJOF2022111638}
T.~{Martinez Ostormujof}, R.~{Purushottam Raj Purohit}, S.~Breumier, N.~Gey, M.~Salib, L.~Germain, Deep learning for automated phase segmentation in ebsd maps. a case study in dual phase steel microstructures, Mater Charact 184 (2022) 111638.
\newblock \href {https://doi.org/10.1016/j.matchar.2021.111638} {\path{doi:10.1016/j.matchar.2021.111638}}.

\bibitem{ronellenfitsch2019inverse}
H.~Ronellenfitsch, N.~Stoop, J.~Yu, A.~Forrow, J.~Dunkel, Inverse design of discrete mechanical metamaterials, Phys Rev Mater 3~(9) (2019) 095201.
\newblock \href {https://doi.org/10.1103/PhysRevMaterials.3.095201} {\path{doi:10.1103/PhysRevMaterials.3.095201}}.

\bibitem{callewaert2016inverse}
F.~Callewaert, S.~Butun, Z.~Li, K.~Aydin, Inverse design of an ultra-compact broadband optical diode based on asymmetric spatial mode conversion, Sci Rep 6~(1) (2016) 1--10.
\newblock \href {https://doi.org/10.1038/srep32577} {\path{doi:10.1038/srep32577}}.

\bibitem{ZENG2023107920}
Q.~Zeng, Z.~Zhao, H.~Lei, P.~Wang, A deep learning approach for inverse design of gradient mechanical metamaterials, Int J Mech Sci 240 (2023) 107920.
\newblock \href {https://doi.org/10.1016/j.ijmecsci.2022.107920} {\path{doi:10.1016/j.ijmecsci.2022.107920}}.

\bibitem{liu2018generative}
Z.~Liu, D.~Zhu, S.~P. Rodrigues, K.-T. Lee, W.~Cai, Generative model for the inverse design of metasurfaces, Nano Lett 18~(10) (2018) 6570--6576.
\newblock \href {https://doi.org/10.1021/acs.nanolett.8b03171} {\path{doi:10.1021/acs.nanolett.8b03171}}.

\bibitem{peurifoy2018nanophotonic}
J.~Peurifoy, Y.~Shen, L.~Jing, Y.~Yang, F.~Cano-Renteria, B.~G. DeLacy, J.~D. Joannopoulos, M.~Tegmark, M.~Solja{\v{c}}i{\'c}, Nanophotonic particle simulation and inverse design using artificial neural networks, Sci Adv 4~(6) (2018) eaar4206.
\newblock \href {https://doi.org/10.1126/sciadv.aar4206} {\path{doi:10.1126/sciadv.aar4206}}.

\bibitem{shiraiwa2022exploration}
T.~Shiraiwa, S.~Kato, F.~Briffod, M.~Enoki, Exploration of outliers in strength-ductility relationship of dual-phase steels, Sci Technol Adv Mater: Methodss 2 (2022) 175--197.
\newblock \href {https://doi.org/10.1080/27660400.2022.2080483} {\path{doi:10.1080/27660400.2022.2080483}}.

\bibitem{hiraide2021application}
K.~Hiraide, K.~Hirayama, K.~Endo, M.~Muramatsu, Application of deep learning to inverse design of phase separation structure in polymer alloy, Comput Mater Sci 190 (2021) 110278.
\newblock \href {https://doi.org/10.1016/j.commatsci.2021.110278} {\path{doi:10.1016/j.commatsci.2021.110278}}.

\bibitem{adachi2020effect}
T.~Adachi, A.~Ito, H.~Adachi, S.~Torizuka, Effect of prior structure to intercritical annealing on rapid formation of ultrafine ferrite+ austenite structure and mechanical properties in 0.1\% c-2\% si-5\% mn steels, ISIJ Int 60~(4) (2020) 764--773.
\newblock \href {https://doi.org/10.2355/isijinternational.ISIJINT-2019-401} {\path{doi:10.2355/isijinternational.ISIJINT-2019-401}}.

\bibitem{wang1997three}
Y.~Wang, A.~Khachaturyan, Three-dimensional field model and computer modeling of martensitic transformations, Acta Mater 45~(2) (1997) 759--773.
\newblock \href {https://doi.org/10.1016/S1359-6454(96)00180-2} {\path{doi:10.1016/S1359-6454(96)00180-2}}.

\bibitem{ALLEN19791085}
S.~M. Allen, J.~W. Cahn, A microscopic theory for antiphase boundary motion and its application to antiphase domain coarsening, Acta Metall 27~(6) (1979) 1085--1095.
\newblock \href {https://doi.org/10.1016/0001-6160(79)90196-2} {\path{doi:10.1016/0001-6160(79)90196-2}}.

\bibitem{khachaturyan2008theory}
A.~G. Khachaturyan, Theory of Structural Transformations in Solids, Courier Corporation, Chelmsford, 2008.

\bibitem{myeong2017effect}
P.~Myeong-Heom, doctral thesis: Effect of Grain Size on Mechanical Properties of Dual Phase Steel Composed of Ferrite and Martensite, Kyoto University, Kyoto, 2017.
\newblock \href {https://doi.org/10.14989/doctor.k20367} {\path{doi:10.14989/doctor.k20367}}.

\bibitem{goodfellow2014generative}
I.~Goodfellow, J.~Pouget-Abadie, M.~Mirza, B.~Xu, D.~Warde-Farley, S.~Ozair, A.~Courville, Y.~Bengio, Generative adversarial nets, Adv Neural Inf Process Syst 27 (2014).
\newblock \href {https://doi.org/10.48550/arXiv.1406.2661} {\path{doi:10.48550/arXiv.1406.2661}}.

\bibitem{goodfellow2014distinguishability}
I.~J. Goodfellow, On distinguishability criteria for estimating generative models (2014).
\newblock \href {https://doi.org/10.48550/arXiv.1412.6515} {\path{doi:10.48550/arXiv.1412.6515}}.

\bibitem{arjovsky2017wasserstein}
M.~Arjovsky, S.~Chintala, L.~Bottou, Wasserstein generative adversarial networks, in: the 34th International Conference on Machine Learning, Vol.~70, 2017, pp. 214--223.
\newblock \href {https://doi.org/10.48550/arXiv.1701.07875} {\path{doi:10.48550/arXiv.1701.07875}}.

\bibitem{knezevic2009crystal}
M.~Knezevic, H.~F. Al-Harbi, S.~R. Kalidindi, Crystal plasticity simulations using discrete fourier transforms, Acta Mater 57~(6) (2009) 1777--1784.
\newblock \href {https://doi.org/10.1016/j.actamat.2008.12.017} {\path{doi:10.1016/j.actamat.2008.12.017}}.

\bibitem{asaro1977strain}
R.~J. Asaro, J.~Rice, Strain localization in ductile single crystals, J Mech Phys Solids 25~(5) (1977) 309--338.
\newblock \href {https://doi.org/10.1016/0022-5096(77)90001-1} {\path{doi:10.1016/0022-5096(77)90001-1}}.

\bibitem{asaro1979geometrical}
R.~J. Asaro, Geometrical effects in the inhomogeneous deformation of ductile single crystals, Acta Mater 27~(3) (1979) 445--453.
\newblock \href {https://doi.org/10.1016/0001-6160(79)90036-1} {\path{doi:10.1016/0001-6160(79)90036-1}}.

\bibitem{hutchinson1976bounds}
J.~W. Hutchinson, Bounds and self-consistent estimates for creep of polycrystalline materials, Proc R Soc Lond A 348~(1652) (1976) 101--127.
\newblock \href {https://doi.org/10.1098/rspa.1976.0027} {\path{doi:10.1098/rspa.1976.0027}}.

\bibitem{pan1983rate}
J.~Pan, J.~R. Rice, Rate sensitivity of plastic flow and implications for yield-surface vertices, Int J Solids Struct 19~(11) (1983) 973--987.
\newblock \href {https://doi.org/10.1016/0020-7683(83)90023-9} {\path{doi:10.1016/0020-7683(83)90023-9}}.

\bibitem{kujirai2020modelling}
S.~Kujirai, K.~Shizawa, Modelling and simulation of dynamic recrystallisation based on multi-phase-field and dislocation-based crystal plasticity models, Philos Mag 100~(16) (2020) 2106--2127.
\newblock \href {https://doi.org/10.1080/14786435.2020.1756501} {\path{doi:10.1080/14786435.2020.1756501}}.

\bibitem{ohashi1994numerical}
T.~Ohashi, Numerical modelling of plastic multislip in metal crystals of fcc type, Philos Mag A 70~(5) (1994) 793--803.
\newblock \href {https://doi.org/10.1080/01418619408242931} {\path{doi:10.1080/01418619408242931}}.

\bibitem{kimura2020crystal}
Y.~Kimura, R.~Ueta, K.~Shizawa, Crystal plasticity fe simulation for kink band formation in mg-based lpso phase using dislocation-based higher-order stress model, Mech Eng J 7~(4) (2020) 19--00612.
\newblock \href {https://doi.org/10.1299/mej.19-00612} {\path{doi:10.1299/mej.19-00612}}.

\bibitem{bailey1960dislocation}
J.~Bailey, P.~Hirsch, The dislocation distribution, flow stress, and stored energy in cold-worked polycrystalline silver, Philos Mag 5~(53) (1960) 485--497.
\newblock \href {https://doi.org/10.1080/14786436008238300} {\path{doi:10.1080/14786436008238300}}.

\bibitem{bishop2006pattern}
C.~M. Bishop, N.~M. Nasrabadi, Pattern recognition and machine learning, Springer, New York, 2006.

\bibitem{goodfellow2016deep}
I.~Goodfellow, Y.~Bengio, A.~Courville, Deep learning, MIT Press, Cambridge, 2016.

\bibitem{tsuji2008managing}
N.~Tsuji, N.~Kamikawa, R.~Ueji, N.~Takata, H.~Koyama, D.~Terada, Managing both strength and ductility in ultrafine grained steels, ISIJ Int 48~(8) (2008) 1114--1121.
\newblock \href {https://doi.org/10.2355/isijinternational.48.1114} {\path{doi:10.2355/isijinternational.48.1114}}.

\bibitem{bessa2017framework}
M.~A. Bessa, R.~Bostanabad, Z.~Liu, A.~Hu, D.~W. Apley, C.~Brinson, W.~Chen, W.~K. Liu, A framework for data-driven analysis of materials under uncertainty: Countering the curse of dimensionality, Comput Methods Appl Mech Eng 320 (2017) 633--667.
\newblock \href {https://doi.org/10.1016/j.cma.2017.03.037} {\path{doi:10.1016/j.cma.2017.03.037}}.

\bibitem{hiraide2023inverse}
K.~Hiraide, Y.~Oya, M.~Suzuki, M.~Muramatsu, Inverse design of polymer alloys using deep learning based on self-consistent field analysis and finite element analysis, Mater Today Commun 37 (2023) 107233.
\newblock \href {https://doi.org/10.1016/j.mtcomm.2023.107233} {\path{doi:10.1016/j.mtcomm.2023.107233}}.

\end{thebibliography}





\end{document}